\documentclass[preprint,floats,tightenlines,11pt,eqsecnum,aps]{revtex4}
\usepackage{amsmath}
\usepackage{graphicx}
\begin{document}
\def\appls{\hbox{$<$\kern-.75em\lower 1.00ex\hbox{$\sim$}}}
\title{$\rho^0(770)-f_0(980)$ MIXING AND $CPT$ VIOLATION\\
IN A NON-UNITARY EVOLUTION OF PION CREATION PROCESS\\ $\pi^- p \to \pi^- \pi^+ n$ ON POLARIZED TARGET}

\author{Miloslav Svec\footnote{electronic address: svec@hep.physics.mcgill.ca}}
\affiliation{Physics Department, Dawson College, Montreal, Quebec, Canada H3Z 1A4}
\date{September 17, 2007}

\begin{abstract}

Unitary evolution from pure initial states to pure final states in $\pi^- p \to \pi^- \pi^+ n$ imposes constraints on pion production amplitudes that are violated by the CERN data on polarized target at 17.2 GeV/c~\cite{svec07a}. The pion creation process is a non-unitary evolution of initial state $\rho_i(\pi^-p)$ to final state $\rho_f(\pi^- \pi^+ n)$ arising from a unitary co-evolution of the pion creation process with a quantum environment described by Kraus representation. The purpose of this work is to identify the interacting degrees of freedom of the environment in a high resolution amplitude analysis of CERN data on $\pi^- p \to \pi^- \pi^+ n$ on polarized target for dipion masses 580-1080 MeV where $S$- and $P$-wave transversity amplitudes dominate. The $S$-wave spectra show presence of $\rho^0(770)$ arising from the presence of $\rho^0(770)$ peak in data component $a_1+a_2$ in relation $|S|^2=a_1+a_2-3|L|^2$ while the $P$-wave spectra $|L|^2$ show a dip at $f_0(980)$ mass arising from $f_0(980)$ structures at 980 MeV in all data. The observed $\rho^0(770)-f_0(980)$ mixing is encoded in all measured density matrix elements which also encode a level splitting of the spectra arising from the interaction of the pion creation process with the environment. The analytical form of the level splitting reveals the existence of a new quantum number $g$ characterizing the environment and allows to identify the four sets of solutions for the transversity amplitudes with the four co-evolution amplitudes required by the Kraus representation. We propose a model for the $CPT$ violating and non-dissipative interaction of the pion creation process with the environment in which non-diagonal transitions between resonant $q \overline {q}$ modes and $\pi^- \pi^+$ states lead to vector-scalar mixing. The model predicts quantum entanglement of $\pi^-\pi^+$ isospin states. The final states in $\pi^- p \to \pi^- \pi^+ n$ do not posses prepareable $CPT$ conjugate states due to entanglement of $\pi^- \pi^+$ pairs and because the environment states do not have well defined charge conjugate states. As a result the concept of $CPT$ symmetry looses its meaning in the unitary co-evolution of the pion creation process with quantum  environment.

\end{abstract}

\maketitle

\tableofcontents

\newpage

\section{Introduction - The puzzle of $\sigma(750)$ resonance.}

Following the discovery in 1961 of $\rho$ meson in $\pi N \to \pi \pi N$ reactions~\cite{erwin61}, the measurements of forward-backward asymmetry in $\pi^- p \to \pi^- \pi^+n$ suggested the existence of a rho-like resonance in the $S$-wave, later referred to as $\sigma(750)$ scalar meson~\cite{hagopian63,islam64,patil64,durand65,baton65}. It was expected that the $\sigma(750)$ resonance would show up prominently in 
$\pi^- p \to \pi^0 \pi^0 n$ production where the $\rho^0(770)$ does not contribute. However the measurements of this reaction at CERN in 1972 found no evidence for a rho-like $\sigma(750)$~\cite{apel72}. Furthermore, in 1973, Pennington and Protopopescu used analyticity and unitarity constraints on partial wave amplitudes in $\pi \pi \to \pi \pi$ scattering (Roy equations) to show that a narrow $\sigma(750)$ resonance cannot contribute to $\pi \pi$ scattering~\cite{pennington73}. From these facts it was concluded that $\sigma(750)$ does not exist and in 1974 Particle Data Group dropped this state from its listings.\\ 

In 1972, van Rossum and his spin physics group at Saclay reported the first measurements of recoil nucleon polarization in $\pi N \to \pi N$ elastic scattering made at CERN at 6 and 16  GeV/c~\cite{lesquen72}. The resulting complete set of observables enabled the first model independent amplitude analysis of a hadronic reaction~\cite{cozzika72}. The results invalidated all Regge models (but not the concept of Regge poles). These findings established that the experimental determination of amplitudes in hadron scattering using measurements with spin is crucial for our understanding of hadron dynamics.\\ 

In 1978, Lutz and Rybicki extended the concept of amplitude analysis to pion production processes. They showed~\cite{lutz78} that almost complete amplitude analysis of reactions $\pi N \to \pi^+ \pi^- N$ and $KN \to K^+ \pi^- N$ is possible from measurements in a single experiment on a transversely polarized target. The work of Lutz and Rybicki opened a new approach to hadron spectroscopy and hadron dynamics by enabling us to study the production of resonances on the level of spin amplitudes rather than spin-averaged cross-sections.\\

The pion production on polarized targets was measured at CERN in several processes. CERN-Munich-Cracow group measured $\pi^- p \to \pi^-\pi^+n$ on transversely polarized proton target in a high statistics experiment at 17.2 GeV/c at low momentum transfers $-t=0.005 - 0.20$ (GeV/c)$^2$~\cite{becker79a,becker79b,chabaud83} and at high momentum transfers $-t=0.20 - 1.00$ (GeV/c)$^2$~\cite{rybicki85}. Saclay group measured $\pi^+ n \to \pi^+ \pi^- p$~\cite{lesquen85,svec92a} and $K^+ n \to K^+ \pi^- p$~\cite{lesquen89,svec92b} on transversely polarized deuteron target at 5.98 and 11.85 GeV/c at larger momentum transfers
$-t = 0.20 - 0.40$ (GeV/c)$^2$. Recently measurements of $\pi^- p \to \pi^-\pi^+n$ on transversely polarized proton target at 1.78 GeV/c at low momentum transfers $-t=0.005 - 0.20$ (GeV/c)$^2$ were made at ITEP~\cite{alekseev99}.\\

The CERN measurements of $\pi^- p \to \pi^-\pi^+n$ and $\pi^+ n \to \pi^+ \pi^- p$ on polarized targets reopened the question of the existence of $\sigma(750)$ scalar meson. Evidence for a narrow $\sigma(750)$ was found in amplitude analyses of $\pi^- p \to \pi^-\pi^+n$ at 17.2 GeV/c~\cite{donohue79,becker79b,rybicki85,svec92c} and in $\pi^+ n \to \pi^+ \pi^- p$ at 5.98 and 11.85~\cite{svec84,svec92c}. Clear evidence for $\sigma(750)$ emerged from later and more precise amplitude analyses of both reactions~\cite{svec96,svec97a,svec02a}. New evidence for $\sigma(750)$ comes from the amplitude analysis of the ITEP data at 1.78 GeV/c~\cite{alekseev99}. The best latest fit in Ref.~\cite{svec02a} gives $m_\sigma=778 \pm 16$ MeV and $\Gamma_\sigma = 142 \pm 33$ MeV. These values are very close to resonance parameters of $\rho(770)$ with $m_\rho = 775.8 \pm 0.5$ MeV and $\Gamma_\rho = 150.3 \pm 1.6$ MeV.\\

In 2001, E852 Collaboration at BNL reported high statistics measurements $\pi^- p \to \pi^0 \pi^0 n$ at 18.3 GeV/c~\cite{gunter01}. The data revealed large differences in the $S$-wave intensities in $\pi^- \pi^+$ and $\pi^0 \pi^0$ production. There was no evidence for a rho-like $\sigma(750)$ resonance in the $\pi^0 \pi^0$ $S$-wave. The BNL data presented anew the puzzle of the $\sigma(750)$ resonance. The CERN and BNL data are both high quality data that cannot be used to exclude one another. We must accept that they are both correct and that the apparent contradictions between them are telling us something new and important. The puzzle of $\sigma(750)$ resonance has become a unique opportunity to learn new physics.\\

The first hint on the solution of the puzzle emerged from the new amplitude analysis of CERN measurements of $\pi^- p \to \pi^- \pi^+ n$ in Ref.~\cite{svec02a}. This study extended the dipion mass range of 600 - 900 MeV of previous analyses~\cite{svec96,svec97a} to 580 - 1080 MeV to include the $f_0(980)$ resonance. Both solutions for the $P$-wave amplitude $|L_d|^2$ showed an unexpected dip at $\sim$ 980 MeV, the mass of $f_0(980)$ resonance. In 2001, Gale and collaborators investigated $\rho(770)-a_0(980)$ mixing in lepton pair production at RHIC arising from the violation of Lorentz symmetry by vacuum and found large measurable effects~\cite{gale01}. Their work suggested that the dip in $|L_d|^2$  arises from $\rho(770)-f_0(980)$ mixing in the $P$-wave. Similar mixing in the $S$-wave then allows to identify $\sigma(750)$ with $\rho^0(770)$ which immediately explains why no $\sigma(750)$ was ever found in $\pi^- p \to \pi^0 \pi^0 n$ production.\\

The $S$-wave intensities in $\pi^+ n \to \pi^+ \pi^- n$ at 11.85 GeV/c and $\pi^- p \to \pi^0 \pi^0$ at 18.3 GeV/c show particularly large differences in their structure for masses below $\sim$ 600 MeV. The expectation that these intensities should be similar in this mass range can be traced to the isospin structure of the amplitudes which originates in the assumption of Generalized Bose-Einstein symmetry for two-pion states. The assumption of Generalized Bose-Einstein symmetry leads to three relations that must be satisfied by combinations of partial wave intensities of $\pi^- \pi^+$, $\pi^0 \pi^0$ and $\pi^+ \pi^+$ production for even dipion spins. In a sequel paper we will show that the relations are all violated by the available data indicating a violation of Generalized Bose-Einstein symmetry in $\pi^- p \to \pi^- \pi^+ n$.\\

There are four sets of physical solutions for the $S$- and $P$- wave transversity amplitudes $A_u(i)$ and $A_d(j), i,j=1,2$ with target nucleon transversity "up" (u) and "down" (d) relative to the scattering plane. All these solutions involve $\rho^0(770)-f_0(980)$ mixing and encode relevant information about the pion creation dynamics. The multitude of the physical solutions may reflect important new aspects of reality. To select one of these solutions as the only valid physical solution amounts to a loss of relevant information. We can retain the full information about the dynamics encoded in the amplitudes from all solutions if we accept that all these solutions are valid physical solutions. But then the measured final state density matrix $\rho_f$ must be a mixed state formed by density matrices $\rho_f(ij)$ with probabilities $p_{ij}$ corresponding to all four solutions $i,j=1,2$
\begin{equation}
\rho_f(\pi^- \pi^+ n)=p_{11}\rho_f(11)+p_{12}\rho_f(12)+p_{21}\rho_f(21)+p_{22}\rho_f(22)
\end{equation}
where $\sum p_{ij}=1$. The final state is a mixed state even when all states $\rho_f(ij)$ are pure states arising from pure initial states. If the hypothesis (1.1) is true then pure initial states would evolve into mixed final states in $\pi^- p \to \pi^- \pi^+ n$ and the dynamics of the pion creation process would be a non-unitary evolution. The non-unitary dynamics could be responsible for the $\rho^0(770)-f_0(980)$ mixing and the violations of Generalized Bose-Einstein symmetry.\\

In 1980 Wald showed that any scattering process of particles that evolves pure initial state into a mixed final state violates $CPT$ symmetry and is time irreversible~\cite{wald80}. In 1982 Hawking showed that in the presence of a black hole (macroscopic or microscopic) a pure initial state of interacting particles will evolve into a mixed final state as some of the quantum states produced in the particle interaction will fall behind the horizon and become inaccessible to measurements by the observer~\cite{hawking82}. He also suggested that quantum fluctuations of the space-time metric will have the same effect on interacting particles and induce their non-unitary evolution - at any energy~\cite{hawking82,hawking84}. Hawking questioned the universal validity of the unitary time evolution in Quantum Field Theory in the presence of metric fluctuations and suggested that initial and final state density matrices $\rho_{in}$ and $\rho_{out}$ are connected by a linear but non-unitary evolution superoperator.\\

Hawking's ideas inspired suggestions to test them experimentally. In 1984, Ellis, Hagelin,
Nanopoulos and Srednicki proposed that quantum fluctuations of the metric form an environment with which interacting as well as free particles interact as open quantum 
systems~\cite{ellis84}. Such interactions would lead to an observable change of coherence and $CPT$ violations in $K^0 \overline{K}^0$ systems. Over the years other suggestions to test Hawking's ideas have been put 
forward~\cite{huet95,ellis96,gerber98,gerber04,mavromatos06,bernabeu06a,bernabeu06b}. Recent experiments with free neutral kaons have yielded remarkably sensitive results on violations of $CPT$ symmetry, time reversal invariance and entanglement of kaon pairs~\cite{fidecaro06,ambrosino06}. So far these experiments did not provide a conclusive confirmation of a non-unitary evolution, and thus possible evidence for quantum gravity effects.\\

In our previous work~\cite{svec07a} we returned to the original Hawking's idea and searched for evidence of a non-unitary evolution of pure initial states of interacting particles into  mixed final states in the existing CERN data on $\pi^- p \to \pi^- \pi^+ n$ on polarized target at 17.2 GeV/c. The purity of the final state $\rho_f$ is controlled by the purity of the recoil nucleon polarization. We developed a spin formalism to calculate the expressions for recoil nucleon polarization for two specific measured initial pure states. Imposing the condition of purity on the recoil nucleon polarization we obtained conditions on the amplitudes which are violated by the model independent amplitude analyses of the CERN data on polarized target at large momentum transfers. We conclude that pure states can evolve into mixed states in $\pi^- p \to \pi^- \pi^+ n$.\\

In quantum theory such non-unitary evolution occurs in open quantum systems or processes $S$ interacting with a quantum environment $E$~\cite{kraus83,nielsen00,breuer02}. The co-evolution of the system $S$ and the environment $E$ is unitary 
\begin{equation}
\rho_f(S,E)=U \rho_i(S) \otimes \rho_i(E)U^+
\end{equation}
The reduced density matrix
\begin{equation}
\rho_f(S)=Tr_E(\rho_f(S,E)) = \mathcal{E}(\rho_i(S))
\end{equation}
is a mixed state and $\mathcal{E}$ describes the non-unitary evolution from $\rho_i(S)$ to $\rho_f(S)$. It is given by Kraus representation~\cite{kraus83,nielsen00,breuer02}
\begin{equation}
\rho_f(S)=\mathcal{E}(\rho_i(S))=\sum \limits_\ell \sum \limits_{m,n} p_{mn} S_{\ell m} \rho_i(S)S^+_{n \ell}
\end{equation}
where $S_{\ell m}=<e_\ell|U|e_m>$ and $|e_\ell>$ are interacting degrees of freedom of the environment. The initial state of the environment has a general form
\begin{equation}
\rho_i(E)=\sum \limits_{m,n} p_{mn}|e_m><e_n|
\end{equation}
We showed in~\cite{svec07a} that Kraus representation leaves invariant the formalism used in data analyses provided the co-evolution with the environment conserves $P$-parity and quantum numbers of the environment. On general grounds~\cite{nielsen00,svec07a} there are four interacting degrees of freedom of the environment in $\pi^- p \to \pi^- \pi^+ n$. The Kraus representation then has a diagonal form
\begin{equation}
\rho_f(S)=\sum \limits_{\ell=1}^4 p_{\ell \ell} S_{\ell \ell} \rho_i(S)S^+_{\ell \ell}= \sum \limits_{\ell=1}^4 p_\ell \rho_f(\ell)
\end{equation}
The measured density matrix elements are redefined to be environment-averaged density matrix elements. The elements that depend explicitely on the environment are predicted to violate certain phase relations. The prediction is in excellent agreement with the CERN data~\cite{svec07a} and validates the view of pion creation processes as an open quantum systems interacting with a quantum environment.\\ 

The question now arises what are the quantum states $|e_\ell>$ of the environment. Experimentally, the final state in $\pi^- p \to \pi^- \pi^+ n$ must be the mixed state $\rho_f(S)$ given by (1.6) which has a form identical to the mixed state $\rho_f(\pi^- \pi^+ n)$ given by (1.1). The Central Hypothesis put forward in this work is that the two states are the same states. The hypothesis then alows us to relate the solutions for the amplitudes with quantum states of the environment.\\

We associate with the two solutions for amplitudes $A_u(i)$ and $A_d(j)$ two qubit states $|i>$ and $|j>$ where $i,j=1,2$ and identify the quantum states $|e_\ell>$ with two-qubit states $|i>|j>$. Experimentally, the two solutions for $S$- and $P$-wave amplitudes $A_u(i)$ and $A_d(j)$ originate in two solutions for mass spectra $|A_u(i)|^2$ and $|A_d(j)|^2$ which can be thought of as a form of level splitting arising from an interaction with the environment. We show in this work that the level splitting is characterized by a new quantum number $g= \pm 1$ associated with the environment. The two solutions for the amplitudes then 
reflect their dependence on this quantum number with $A_u(i)=A_u(g_u)$ and $A_d(j)=A_u(g_d)$.
The quantum states $|e_\ell>$ are then identified with two-qubit states $|g_u g_d>$.\\

The new quantum number $g$ of the environment gives rise to a new $CPT$ violating interaction resposible for the level splitting and the non-unitary evolution in $\pi^- p \to \pi^- \pi^+ n$. We formulate a model of such interaction in which the environment states $|g_ug_d>$ interact with a coherent state of $q \overline{q}$ resonant modes produced in an intermediate stage $\pi^- p \to q \overline{q} n$ . The model explains the observed $\rho^0(770)-f_0(980)$ mixing in terms of environment induced transitions of $q \overline{q}$ resonant modes with a definite spin $K$ to a superposition of two-pion states with different spins $J$. The model leads naturally to a violation of Generalized Bose-Einstein symmetry and to dynamic entanglement of $\pi^- \pi^+$ pairs with definite spin $J$.\\

The interaction of pion creation process with the environment can be thought of as a scattering of an intermediate coherent state of $q \overline{q}$ resonant modes carrying energy-momentum and spins with particles of the environment carrying quantum entanglement (1.5). There is no exchange of energy-momentun between the hadrons and the environment in this non-dissipative, entanglement changing interaction. The observed process is time-irreversible and violates $CPT$ symmetry.\\

In this work we focus on the $S$- and $P$-wave decoherence free subspace of the reduced density matrix $\rho_f^0$ which is the trace over the unmeasured recoil nucleon spin states.
The density matrix elements of a decoherence free subspace do not depend explicitely on the interaction with the environment. In Section II. we show that all $S$- and $P$-wave elements of $\rho_f^0$ are expressed in terms of reduced transversity amplitudes which do not depend on the relative phase $\omega$ between the $S$-wave transversity amplitudes $S_u$ and $S_d$. Phase relations arising from the assumption of decoherence free subspace result in cubic equations for $P$-wave amplitudes $|L_u|^2$ and $|L_d|^2$ which render the system analytically solvable from data on matrix elements $Re \rho^0_u$ and $Re \rho^0_y$ measured with transversely polarized targets.\\

In Section III. we present results of high resolution amplitude analysis of CERN data on $\pi^- p \to \pi^- \pi^+ n$ at 17.2 GeV/c for dipion masses 580-1080 MeV as one of 8 reactions analyzed. The analysis used 1 and 5 million Monte Carlo samplings of the data error volume in each mass bin. The analysis with 5 million samplings gives essentially identical results, indicating the resolution of mass spectra $|A_\tau|^2, \tau=u,d$ is stable. The relative phases are determined up to a sign ambiguity and a relative phase $\omega$ between $S$-wave amplitudes $S_d$ and $S_u$. In Section IV. we relate the positivity of density matrix and decoherence assumption to observation that the relative phases between $S_\tau$ and $L_\tau$ amplitudes do not change sign which allows the unique asignment of signs to the relative phases. In Section V. we show how the measurements of matrix elements $Im \rho^0_x$ and $Im \rho^0_z$ resolves the sign ambiguity of the relative phases. In a sequel paper~\cite{svec07c} we show that the phase $\omega$ is uniquely determined by a conversion of transversity amplitudes into helicity amplitudes. As a result there are four sets of amplitudes $A_u(i)$ and $A_d(j)$ consistent with the requirement of Kraus representation (1.6).\\

In Section VI. we present evidence for $\rho^0(770)-f_0(980)$ mixing. Resonace peaks of $\rho^0(770)$ and $f_0(980)$ are clearly resolved in $S$-wave amplitudes $|S_d|^2$ in contrast to our previous low resolution analyses using 40 000 
samplings ~\cite{svec96,svec97a,svec02a}. The $S$-wave moduli are given by a relation 
$|S|^2=a_1+a_2-3|L|^2$ where $a_1+a_2$ is a data component. The presence of $\rho^0(770)$ in $S$-wave spectra $|S|^2$ arises from the presence of $\rho^0(770)$ peak in the data component $a_1+a_2$ which survives the subtraction of $\rho^0(770)$ peak in the $P$-wave amplitude $3|L|^2$. In Section VII. we present a test of rotational and Lorentz symmetry in $\pi^- p \to \pi^- \pi^+ n$. We show that the large differences observed in $\rho^0(770)$ widths in unnatural and natural exchange amplitudes $|U_d|^2$ and $|N_d|^2$ arise from interference of $P$-wave amplitudes with helicities $\lambda=+1$ and $-1$ and not from a violation of Lorentz symmetry - which if true could explain also the $\rho^0(770)-f_0(980)$ mixing.\\

In Section VIII. we show that the data exclude the existence of a single physical solution for the moduli. Instead the data require that the moduli $|A_\tau|^2$ have a specific analytical form of two roots of the cubic equation which can be interpreted as a level splitting of the spectra. In Section IX. we show that the two solutions for the moduli $|A_\tau|^2$ can be rewritten as interferences of two amplitudes $Z_{A,\tau}(g_\tau)$ with phases $g_\tau \lambda_{A, \tau}$ characterized by a new quantum number $g_\tau=\pm1$. The phase $\lambda_{A,\tau}$ determines the level splitting. The amplitudes $A_u(i)$ and $A_d(j)$, $i,j=1,2$ correspond to amplitudes $A_u(g_u)$ and $A_d(g_d)$ with $g_u,g_d= \pm1$. In Section X. we identify the quantum states of the environment with two-qubit states $|i>|j>$, or alternalively, with $|g_u>|g_d>$ and define co-evolution amplitudes of $\pi^- p \to \pi^- \pi^+ n$ process with the environment.\\

In Section XI. we formulate a model of interaction of pion creation process with the environment to explain $\rho^0(770)-f_0(980)$ mixing. In the first stage of pion creation process a coherent state is formed of resonant $q \overline{q}$ modes with spins $K(q \overline {q})$ and isospins $I_K$. The interaction of the coherent state with the environment results in transitions of resonant modes with spin $K$ and isospin $I_K$ to two-pion states with spins $J(\pi^-\pi^+)$ which gives rise to $\rho_0(770)-f_0(980)$ mixing. In Section XII. we introduce Generalized Bose-Einstein symmetry conserving and violating co-evolution amplitudes corresponding to maximally entangled $\pi^- \pi^+$ states. The total co-evolution amplitude is their superposition which leads to dynamic entanglement of the observed $\pi^- \pi^+$ states. The change in the entanglement content of the $\pi^- \pi^+$ states arises entirely from the $\rho^0(770)-f_0(980)$ mixing which in turn originates in the $CPT$ violating non-local interaction with the environment. The paper closes with a brief summary in Section XIII.

\newpage

\section{$S$- and $P$-wave decoherence free subspace of reduced density \\ matrix in $\pi^- p \to \pi^- \pi^+ n$ and similar processes.}

In previous paper~\cite{svec07a} we have developed a general spin formalism for the final state density matrix in $\pi^- p \to \pi^- \pi^+ n$ and similar pion creation processes, and defined nucleon helicity and transversity amplitudes with definite $t$-channel naturality. In the following we assume a familiarity with the Section II. of this paper.\\

The most feasible experiments are measurements of $\pi^- p \to \pi^- \pi^+ n$ on unpolarized or polarized targets with target polarization $\vec {P}=(P_x,P_y,P_z)$. In modern polarized targets the direction of the  polarization vector $\vec {P}$ can be selected at will. The experiments measure the two-pion angular distribution $I^0(\theta \phi, \vec{P})$ leaving the recoil nucleon polarization vector $\vec{Q}$ not observed. Such measurements provide information on the reduced final state density matrix $I^0(\theta \phi,\vec{P})$ given 
by~\cite{svec07a} 
\begin{equation}
I^0(\theta \phi,\vec{P}) = Tr_{\chi = \chi'} ((\rho_f(\theta \phi,\vec{P})^{{1 \over{2}} 
{1 \over{2}}}_{\chi \chi'})=I^0_u(\theta \phi)+P_x I^0_x(\theta \phi) + P_y I^0_y(\theta \phi) + P_z I^0_z(\theta \phi)
\end{equation}
where the components of $I^0(\theta \phi,\vec{P})$ are
\begin{equation}
I^0_k(\theta \phi) = {{d^2 \sigma} \over{dtdm}}
\sum \limits_{J \leq J'}^{J_{max}} \sum \limits _{\lambda \geq 0} \sum \limits_{\lambda'} 
\xi_{JJ'} \xi_\lambda (Re \rho^0_k)^{JJ}_{\lambda \lambda'} Re(Y^J_\lambda(\theta\phi)Y^{J*}_{\lambda'}(\theta \phi))
\end{equation}
for k=u (unpolarized target) and k=y (transversely polarized taget normal to the scattering plane), and
\begin{equation}
I^0_k(\theta \phi) = {{d^2 \sigma} \over{dtdm}} 
\sum \limits_{ J \leq J'}^{J_{max}} \sum \limits_{\lambda \geq 0} \sum \limits_{\lambda'} 
\xi_{JJ'} \xi_\lambda (Im \rho^0_k)^{JJ'}_{\lambda \lambda'} 
Im(Y^J_\lambda(\theta \phi)Y^{J'*}_{\lambda'}(\theta \phi))
\end{equation}
for k=x (transversely polarized target in the scattering plane) and k=z (longitudinally polarized target). In (2.2) and (2.3) $\xi_0=1$ and $\xi_\lambda=2$ for $\lambda >0$ and the factor $\xi_{JJ'}=1$ for $J=J'$ and $\xi_{JJ'}=2$ for $J<J'$. Experimentally, in a given region of dimeson mass $m$ and momentum transfer $t$ only amplitudes with $J \leq J_{max}$ contribute and all sums in (2.2) and (2.3) are finite. It is the intensity $I^0(\theta \phi, \vec{P})$ which has been measured in CERN measurements of pion creation processes on transversely polarized targets and from which $S$- and $P$-wave density matrix elements were determined in small $(m,t)$ bins using maximum likelihood method~\cite{eadie71,grayer74,becker79a,lesquen85}.\\

General expressions for density matrix elements in terms of amplitudes were tabulated by Lutz and Rybicki~\cite{lutz78} and are reproduced in Appendix A of Ref.~\cite{svec07a}. For $S$- and $P$-wave amplitudes we shall use a simplified notation $A_\tau$ where $A=S,L,U,N$
are transversity amplitudes with definite $t$-channel naturality
\begin{equation}
S_\tau=U^0_{0,\tau}, \quad L_\tau=U^1_{0,\tau}
\end{equation}
\[
U_\tau=U^1_{1,\tau}, \quad N_\tau=N^1_{1,\tau}
\]
In (2.4) $U^J_{\lambda, \tau}$ and $N^J_{\lambda, \tau}$ are unnatural and natural exchange transversity amplitudes introduced in Ref.~\cite{svec07a} and $J$ and $\lambda$ are dimeson spin and helicity.  The nucleon transversity $\tau=u,d$ for target spin "up" or "down" relative to the scattering plane, respectively. In $\pi^- p \to \pi^- \pi^+n$ the amplitudes $U^J_{\lambda,\tau}$ exchange $\pi$ and $a_1$ quantum numbers in the $t$-channel while the amplitudes $N^J_{\lambda, \tau}$ exchange $a_2$ quantum numbers.\\

Amplitude analysis of the $S$- and $P$-wave subsystem of the reduced density matrix determines reduced transversity amplitudes defined as follows
\newpage
\begin{equation}
S=|S_u|, \quad \overline {S} = |S_d|
\end{equation}
\[
L=|L_u| \exp i \left (\Phi_{L_u}-\Phi_{S_u} \right), \quad
\overline {L}=|L_d| \exp i \left ( \Phi_{L_d}-\Phi_{S_d} \right )
\]
\[
U=|U_u| \exp i \left (\Phi_{U_u}-\Phi_{S_u} \right). \quad
\overline {U}=|U_d| \exp i \left ( \Phi_{U_d}-\Phi_{S_d} \right )
\]
\[
N=|N_u| \exp i \left (\Phi_{N_u}-\Phi_{S_d} \right), \quad
\overline {N}=|N_d| i \exp \left ( \Phi_{N_d}-\Phi_{S_u} \right )
\]
where $\Phi_{A_\tau}$ is the phase of the amplitude $A_\tau$. The reduced transversity amplitudes are related to transversity amplitudes by phase factors
\begin{equation}
A_u=A \exp {i \Phi_{S_u}}, \qquad A_d=\overline {A} \exp {i \omega} \exp {i \Phi_{S_u}}
\end{equation}
for unnatural exchange amplitudes $A=S,L,U$ and  
\begin{equation}
N_u=N \exp {i \omega} \exp {i \Phi_{S_u}}, \qquad N_d=\overline {N} \exp {i \Phi_{S_u}}
\end{equation}
for natural exchange amplitude $N$. In (2.6) and (2.7) $\Phi_{S_u}$ is the arbitrary absolute phase and $\omega=\Phi_{S_d}-\Phi_{S_u}$ is the relative phase between $S$-wave amplitudes of opposite transversity.\\

\begin{table}
\caption{Density matrix elements $Re \rho^0_u {{d^2 \sigma}/{dtdm}}$ and $Re \rho^0_y {{d^2 \sigma}/{dtdm}}$ in terms of reduced transversity amplitudes.}
\begin{tabular}{ccc}
\toprule
$\rho^{JJ'}_{\lambda \lambda}$&$Re \rho^0_u {{d^2 \sigma}/{dtdm}}$&$Re \rho^0_y {{d^2 \sigma} /{dtdm}}$\\
\colrule
$\rho^{00}_{ss}$&$|S|^2+|\bar {S}|^2$&$|S|^2-|\bar {S}|^2$\\
$\rho^{11}_{00}$&$|L|^2+|\bar {L}|^2$&$|L|^2-|\bar {L}|^2$\\
$\rho^{11}_{11}$&${1\over{2}}(|U|^2+|\bar {U}|^2)+{1\over{2}}(|N|^2+|\bar {N}|^2)$&
${1\over{2}}(|U|^2-|\bar {U}|^2)+{1\over{2}}(|N|^2-|\bar {N}|^2)$\\
$\rho^{11}_{1-1}$&$-{1\over{2}}(|U|^2+|\bar {U}|^2)+{1\over{2}}(|N|^2+|\bar {N}|^2)$&
$-{1\over{2}}(|U|^2-|\bar {U}|^2)+{1\over{2}}(|N|^2-|\bar {N}|^2)$\\
$Re \rho^{10}_{0s}$&$Re (L S^*+\bar {L} \bar {S}^*)$&$Re (LS^*-\bar {L} \bar {S}^*)$\\
$\sqrt{2}Re \rho^{10}_{1s}$&$Re (US^*+\bar {U} \bar {S}^*)$&$Re (US^*-\bar {U} \bar {S}^*)$\\
$\sqrt{2}Re \rho^{11}_{01}$&$Re (LU^*+\bar {L} \bar {U}^*)$&$Re (LU^*-\bar {L} \bar {U}^*)$\\
\botrule
\end{tabular}
\label{Table I.}
\end{table}

\begin{table}
\caption{Density matrix elements $Im \rho^0_x {{d^2 \sigma}/{dtdm}}$ and $Im \rho^0_z {{d^2 \sigma}/{dtdm}}$ in terms of reduced transversity amplitudes.}
\begin{tabular}{ccc}
\toprule
$\rho^{JJ'}_{\lambda \lambda}$&$Im \rho^0_x {{d^2 \sigma}/{dtdm}}$&$Im \rho^0_z {{d^2 \sigma} /{dtdm}}$\\
\colrule
$\sqrt{2}Im \rho^{01}_{s1}$&$Re(-S \bar {N}^*+N \bar {S}^*)$&$Im(+S \bar {N}^*-N \bar {S}^*)$\\
$\sqrt{2}Im \rho^{11}_{01}$&$Re(-L \bar {N}^*+N \bar {L}^*)$&$Im(+L \bar {N}^*-N \bar {L}^*)$\\
$Im \rho^{11}_{-11}$&$Re(+U \bar {N}^*-N \bar {U}^*)$&$Im(-U \bar {N}^*+N \bar {S}^*)$\\
\botrule
\end{tabular}
\label{Table II.}
\end{table}

In Tables I. and II. we present $S$- and $P$-wave density matrix elements expressed in terms of the reduced transversity amplitudes. Because of the angular properties of $Y^1_\lambda(\theta \phi)$, the elements $(\rho^0_k)^{00}_{00} \equiv (\rho^0_k)^{00}_{ss}$, $(\rho^0_k)^{11}_{00}$ and $(\rho^0_k)^{11}_{11}, k=u,y$ are not independent but appear as two independent combinations in the measured angular distributions (2.2)
\begin{equation}
(\rho^0_k)_{SP} \equiv (\rho^0_k)^{00}_{ss}+(\rho^0_k)^{11}_{00}+2(\rho^0_k)^{11}_{11}, 
\qquad
(\rho^0_k)_{PP} \equiv (\rho^0_k)^{11}_{00}-(\rho^0_k)^{11}_{11}
\end{equation}
In terms of reduced transversity amplitudes they read
\begin{eqnarray}
(\rho^0_u)_{SP}{{d^2 \sigma} \over{dtdm}} & = &
|S|^2+|\overline {S}|^2+|L|^2+|\overline {L}|^2+|U|^2+|\overline {U}|^2+
|N|^2+|\overline {N}|^2\\
\nonumber
(\rho^0_y)_{SP}{{d^2 \sigma} \over{dtdm}} & = &
|S|^2-|\overline {S}|^2+|L|^2-|\overline {L}|^2+|U|^2-|\overline {U}|^2+
|N|^2-|\overline {N}|^2
\end{eqnarray}
\begin{eqnarray}
(\rho^0_u)_{PP}{{d^2 \sigma} \over{dtdm}} & = &
|L|^2+|\overline {L}|^2-{1\over{2}}(|U|^2+|\overline {U}|^2+|N|^2+|\overline {N}|^2)\\
\nonumber
(\rho^0_y)_{PP}{{d^2 \sigma} \over{dtdm}} & = &
|L|^2-|\overline {L}|^2-{1\over{2}}(|U|^2-|\overline {U}|^2+|N|^2-|\overline {N}|^2)
\end{eqnarray}
Note that in (2.3) $Im(\rho^0_k)^{JJ}_{\lambda \lambda}=0$ and $Im(Y^J_0(\theta \phi)Y^{J'*}_0(\theta \phi))=0$. Also note that elements $Im (\rho^0_k)^{10}_{0s}$, $Im (\rho^0_k)^{10}_{1s}$, $Im (\rho^0_k)^{10}_{10}$, $k=u,y$ and $Re (\rho^0_k)^{01}_{s1}$,
$Re (\rho^0_k)^{11}_{01}$, $Re (\rho^0_k)^{11}_{-11}$, $k=x,z$ are not zero. They are not observable as the result of parity conservation but they are calculable from amplitude analysis.\\

The observables measured in $\pi^- p \to \pi^- \pi^+ n$ on transversely polarized target  organize themselves into two groups involving amplitudes of opposite transversity. Using the expressions in Table I. and $\Sigma={{d^2 \sigma}/{dtdm}}$, the two groups are
\begin{eqnarray}
a_1 & = &{1 \over{2}}((\rho^0_u)_{SP}+(\rho^0_y)_{SP}) \Sigma=|S|^2+|L|^2+|U|^2+|N|^2\\
\nonumber
a_2 & = &((\rho^0_u)_{PP}+(\rho^0_y)_{PP}) \Sigma=2|L|^2 -|U|^2-|N|^2\\
\nonumber
a_3 &= &((\rho^0_u)^{11}_{1-1}+(\rho^0_y)^{11}_{1-1}) \Sigma=|N|^2-|U|^2\\
\nonumber
a_4 &= &{1 \over {2}}((\rho^0_u)^{10}_{0s}+(\rho^0_y)^{10}_{0s}) \Sigma=
|L||S| \cos(\Phi_{LS})\\
\nonumber
a_5 &= &{1 \over {\sqrt{2}}}((\rho^0_u)^{10}_{1s}+(\rho^0_y)^{10}_{1s}) \Sigma=
|U||S| \cos(\Phi_{US})\\
\nonumber
a_6 & = &{1 \over {\sqrt{2}}}((\rho^0_u)^{11}_{01}+(\rho^0_y)^{11}_{01}) \Sigma=
|L||U| \cos(\Phi_{LU})
\end{eqnarray}
for reduced transversity amplitudes with transversity $\tau=u$, and
\begin{eqnarray}
\overline {a}_1 & = &{1 \over{2}}((\rho^0_u)_{SP}-(\rho^0_y)_{SP}) \Sigma=
|\overline {S}|^2+|\overline {L}|^2+|\overline {U}|^2+|\overline {N}|^2\\
\nonumber
\overline {a}_2 & = &((\rho^0_u)_{PP}-(\rho^0_y)_{PP}) \Sigma=
2|\overline {L}|^2 -|\overline {U}|^2-|\overline {N}|^2\\
\nonumber
\overline {a}_3 & = &((\rho^0_u)^{11}_{1-1}-(\rho^0_y)^{11}_{1-1}) \Sigma=
|\overline {N}|^2-|\overline {U}|^2\\
\nonumber
\overline {a}_4 & = &{1 \over {2}}((\rho^0_u)^{10}_{0s}-(\rho^0_y)^{10}_{0s}) \Sigma=
|\overline {L}||\overline {S}| \cos(\overline {\Phi}_{LS})\\
\nonumber
\overline {a}_5 & = &{1 \over {\sqrt{2}}}((\rho^0_u)^{10}_{1s}-(\rho^0_y)^{10}_{1s}) \Sigma=
|\overline {U}||\overline {S}| \cos(\overline {\Phi}_{US})\\
\nonumber
\overline {a}_6 & = &{1 \over {\sqrt{2}}}((\rho^0_u)^{11}_{01}-(\rho^0_y)^{11}_{01}) \Sigma=
|\overline {L}||\overline {U}| \cos(\overline {\Phi}_{LU})
\end{eqnarray}
for reduced transversity amplitudes with transversity $\tau=d$. The relative phases are 
defined as in (2.5). For dipion masses where $S$- and $P$-wave dominate, $(\rho^0_u)_{SP}$ and $(\rho^0_y)_{SP}$ are traces
\begin{equation}
(\rho^0_u)_{SP}=Tr((\rho^0_u)^{JJ}_{\lambda \lambda})=1, \quad
(\rho^0_y)_{SP}=Tr((\rho^0_y)^{JJ}_{\lambda \lambda})=T
\end{equation}
where $T$ is target spin asymmetry~\cite{svec07a}.\\

The 6 equations (2.11) and (2.12) each involve 7 unknowns for 4 moduli and 3 cosines, and are not solvable. The missing equation in each group is supplied not by the data but by phase relations
\begin{equation}
\Phi_{LS}-\Phi_{US}-\Phi_{LU}=
(\Phi_{L_u}-\Phi_{S_u})-(\Phi_{U_u}-\Phi_{S_u})-(\Phi_{L_u}-\Phi_{U_u})=0
\end{equation}
\[
\overline {\Phi}_{LS}-\overline {\Phi}_{US}-\overline {\Phi}_{LU}
=(\Phi_{L_d}-\Phi_{S_d})-(\Phi_{U_d}-\Phi_{S_d})-(\Phi_{L_d}-\Phi_{U_d})=0
\]
These conditions lead to non-linear relations between the cosines
\begin{equation}
\cos^2(\Phi_{LS})+\cos^2(\Phi_{US})+\cos^2(\Phi_{LU})-
2\cos(\Phi_{LS})\cos(\Phi_{US})\cos(\Phi_{LU})=1
\end{equation}
\[
\cos^2(\overline {\Phi}_{LS})+\cos^2(\overline {\Phi}_{US})+\cos^2(\overline {\Phi}_{LU})-
2\cos(\overline {\Phi}_{LS})\cos(\overline {\Phi}_{US})\cos(\overline {\Phi}_{LU})=1
\]
Similar relations also hold for the sines. Substituting from (2.11) and (2.12) we get
\begin{equation}
a_6^2|S|^2+a_5^2|L|^2+a_4^2|U|^2-|S|^2|L|^2|U|^2=2a_4a_5a_6
\end{equation}
\[
\overline{a}_6^2|\overline {S}|^2+\overline {a}_5^2|\overline {L}|^2
+\overline {a}_4^2|\overline {U}|^2-|\overline {S}|^2|\overline {L}|^2|\overline {U}|^2
=2\overline {a}_4\overline {a}_5\overline {a}_6
\]
From (2.11) we have 3 equations for moduli
\begin{eqnarray}
|S|^2 = (a_1+a_2)-3|L|^2\\
\nonumber
|U|^2=|L|^2-{1 \over {2}}(a_2+a_3)\\
\nonumber
|N|^2=|L|^2-{1 \over {2}}(a_2-a_3)
\end{eqnarray}
Substituting (2.17) into the first equation in (2.16) we get a cubic equation for $|L|^2 \equiv x$
\begin{equation}
ax^3+bx^2+cx+d=0
\end{equation}  where $a=3$ and 
\begin{eqnarray}
b & = & -3[{1 \over {3}}(a_1+a_2)+{1 \over {2}}(a_2+a_3)]\\
\nonumber
c & = &{1 \over {2}}(a_1+a_2)(a_2+a_3)+a_4^2+a_5^2-3a_6^2\\
\nonumber
d & = &(a_1+a_2)a_6^2-{1 \over {2}}(a_2+a_3)a_4^2-2a_4a_5a_6
\end{eqnarray}
From (2.12) we get equations for moduli similar to (2.17), and from the second equation in (2.16) we obtain a cubic equation for $|\overline {L}|^2 = \overline {x}$ with coefficients given the expressions (2.18) with observables $\overline {a}_k,k=1,6$ replacing $a_k,k=1,6$.\\

Apart from some factors ${1 \over {2}}$ and ${1 \over{\sqrt{2}}}$, the observables $a_k$ and $\overline {a}_k, k=1,6$ defined in (2.11) and (2.12) are density matrix elements $Re \rho^0_u + Re \rho^0_y$ and $Re \rho^0_u - Re \rho^0_y$ corresponding to pure initial states with target polarization $P_y=+1$ and $P_y=-1$, respectively. The assumptions (2.14) mean that the density matrix elements $a_k, k=1,6$ and $\overline {a}_k,k=1,6$ each form a decoherence free subspace of the reduced density matrix. As the result of a symmetry of the interaction of the pion creation process with the environment, these elements decouple from the environment. The equations (2.11) and (2.12) will thus not be solved by environment-averaged moduli and cosines but separately by environment-dependent moduli and cosines defined by Kraus representation~\cite{svec07a}. On general grounds, there are four interacting degrees of freedom in the interaction of pion creation process $\pi^- p \to \pi^- \pi^+ n$ with the environment leading to four sets of moduli and cosines~\cite{svec07a,nielsen00}\\

For any $J_{max} \geq 1$ the $S$- and $P$- wave density matrix elements form a completely autonomous subsystem of the reduced density matrix in a sense that it is analytically solvable and determines completely the $S$- and $P$-wave transversity amplitudes $A_\tau, \tau=u,d$ as well as helicity amplitudes $A_n,n=0,1$ with definite naturality~\cite{svec07c}. The $S$- and $P$- wave density matrix elements of the reduced density matrix do not depend on the relative phase $\omega$ and involve only the reduced transversity amplitudes (2.5). The phase $\omega$ is determined analytically in the process of conversion of transversity amplitudes $A_\tau, \tau=u,d$ into helicity amplitudes $A_n$ with $n=0,1$ nucleon helicity non-flip and flip, respectively~\cite{svec07c}. As the result the transversity amplitudes (2.6) and (2.7) are known up to an overall absolute phase. In the next Section III. we show that there are four sets $A_u(i),A_d(j),i,j=1,2$ of transversity amplitudes solving the system. We shall identify these four solutions for $S$- and $P$-wave amplitudes with the four environment-dependent $S$- and $P$-wave amplitudes predicted from the Kraus representation describing the co-evolution of the pion creation process with the quantum environment.\\

\section{Determination of reduced $S$- and $P$-wave transversity amplitudes from measurements of $Re \rho^0_u$ and $Re \rho^0_y$.}

The cubic equations for $|L|^2$ and $|\overline {L}|^2$ are just another form of the phase conditions (2.14). The equations can be solved analyticaly. To solve for $x$ we write $x=y-y_0$ and require that the cubic equation $ax^3+bx^2+cx+d=0$ transforms to the form
\begin{equation}
y^3+3Py+2Q=0
\end{equation}
This is accomplished with $y_0=b/3a$ and
\begin{equation}
P=-y_0^2 + \Bigl ({c \over {3a}} \Bigr ), \qquad 
2Q= -y_0 \Bigl (2P+{c \over {3a}} \Bigr ) +{d \over {a}} 
\end{equation}
Next we define quantities
\begin{equation}
R=\text {sign}(Q) \sqrt{|P|}, \qquad V={Q \over {R^3}} \geq 0
\end{equation}
There are three categories of of solutions of cubic equation (3.1)~\cite{bronshtein54}. They are given in Table III. \\

\begin{table}
\caption{Three categories of solutions of Eq. (3.1).}
\begin{tabular}{ccc}
\toprule
$P<0$ & $P<0$ & $P>0$\\
$Q^2+P^3 \leq 0$ & $Q^2+P^3 > 0$ &   \\
$V=\cos(\phi)$ & $V=\cosh(\phi)$ & $V=\sinh(\phi)$ \\
\colrule
$r_1=   +2R \cos  \Bigl ( {{\pi-\phi} \over{3}} \Bigr )$ &
$r_1^{'} =R \cosh \Bigl ({\phi \over {3}} \Bigr )+i\sqrt{3}R \sinh \Bigl ({\phi \over {3}} \Bigr )$ &
$r_1^{''}=R \sinh \Bigl ({\phi \over {3}} \Bigr )+i\sqrt{3}R \cosh \Bigl ({\phi \over {3}} \Bigr )$ \\
$r_2=+2R\cos \Bigl ( {{\pi+\phi} \over{3}} \Bigr )$ & 
$r_2^{'} =r_1^{'*} $ & 
$r_2^{''}=r_1^{''*}$\\
$r_3=     -2R \cos  \Bigl ({\phi \over {3}} \Bigr )$ &
$r_3^{'}= -2R \cosh \Bigl ({\phi \over {3}} \Bigr )$ & 
$r_3^{''}=-2R \sinh \Bigl ({\phi \over {3}} \Bigr )$ \\
\botrule
\end{tabular}
\label{Table IV.}
\end{table}

\begin{table}
\caption{Range of momentum transfer $-t$ and dimeson mass $m$ of CERN data on polarized target for which high resolution amplitude analyses were performed.}
\begin{tabular}{cccccc}
\toprule
No. & Reaction &  $p_{lab}$  &  $-t$  &  $m$  & References \\
    &            & $(GeV/c)$ & $(GeV/c)^2$ & $(MeV)$ &   \\
\colrule
1 & $\pi^- p \to \pi^- \pi^+ n$ & 17.2  & 0.005-0.20  & 580-1080 & \cite{chabaud83,rybicki96}\\
2 & $\pi^- p \to \pi^- \pi^+ n$ & 17.2  & 0.00-1.00  & 710- 830 & \cite{becker79a,degroot79b}\\
3 & $\pi^+ n \to \pi^+ \pi^- p$ & 11.85 & 0.20-0.30  & 360-1040 & \cite{lesquen82,lesquen85}\\
4 & $\pi^+ n \to \pi^+ \pi^- p$ & 11.85 & 0.10-1.00  & 710- 830 & \cite{lesquen82,lesquen85}\\
5 & $\pi^+ n \to \pi^+ \pi^- p$ & 5.98  & 0.20-0.30  & 360-1040 & \cite{lesquen82,lesquen85}\\
6 & $\pi^+ n \to \pi^+ \pi^- p$ & 5.98  & 0.10-1.00  & 710- 830 & \cite{lesquen82,lesquen85}\\
7 & $K^+ n \to K^+ \pi^- p$     & 5.98  & 0.20-0.30  & 812- 972 & \cite{lesquen82,lesquen89}\\
8 & $K^+ n \to K^+ \pi^- p$     & 5.98  & 0.10-1.00  & 842- 942 & \cite{lesquen82,lesquen89}\\
\botrule
\end{tabular}
\label{Table IV.}
\end{table}

Previous analyses of CERN data on $\pi^- p \to \pi^- \pi^+n$, $\pi^+n \to \pi^+ \pi^- p$ and $K^+ n \to K^+ \pi^- p$ in 8 kinematic regions~\cite{svec92a,svec92c,svec92b} which included the study of the complex solutions found in most $(m,t)$ bins $R>0$ and negative values for the third solution $x_3=r_3-y_0$, irrespective of the category of the solution. This allowed later studies~\cite{svec96,svec97a,svec02a} to focus on the category with three real solutions. The focus of the present analysis is also on this category.\\

We have performed high resolution amplitude analysis of CERN data on $\pi^- p \to \pi^- \pi^+n$, $\pi^+n \to \pi^+ \pi^- p$ and $K^+ n \to K^+ \pi^- p$ in 8 kinematic regions listed in Table IV. The analyses used 1 million of Monte Carlo samplings of error volumes of measured density matrix elements. Not all values of density matrix elements correspond to physical amplitudes. The procedure produces frequency distribution of physical solutions for the moduli and phases and for various observables calculated from the solutions for amplitudes. The average values of the distributions for the moduli add up exactly to $\Sigma={{d^2 \sigma}/{dtdm}}$ and those for the phases satisfy cosine conditions (2.15). These averaged values can thus be interpreted as the measured amplitudes. The range of distributions determines the asymmetric error bars on the amplitudes and calculated observables. Analyses of $\pi^- p \to \pi^- \pi^+ n$ using 4 million and 5 million Monte Carlo samplings produce virtually identical averaged values and slightly larger errors, indicating the analysis is stable. \\

The initial aim of the new study was to resolve resonant strucures in the mass spectra in $\pi^- p \to \pi^- \pi^+ n$ at 17.2 GeV/c in the mass range 580-1080 MeV at small momentum transfers $-t$=0.005-0.20 (GeV/c)$^2$. The results for the moduli squared of transversity amplitudes in $\pi^- p \to \pi^- \pi^+ n$ are shown in Figures 1-4. We discuss these results and the evidence they present for $\rho^0(770)-f_0(980)$ mixing in Section VI.. where we also present additional supporting evidence from analyses of several other reactions in Table IV..\\

\begin{figure}[htp]
\includegraphics[width=12cm,height=10.5cm]{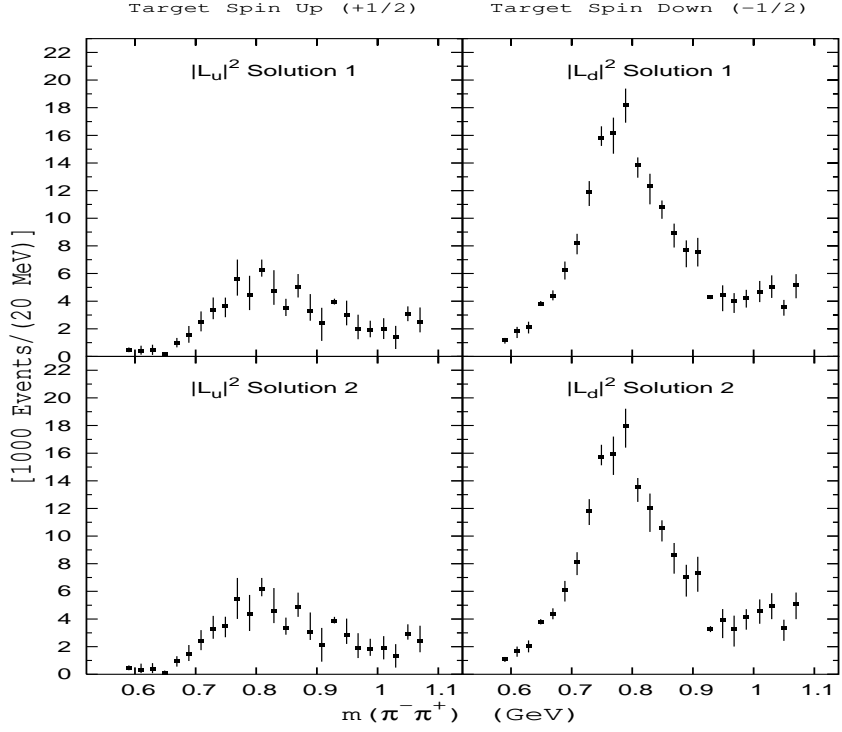}
\caption{Solutions 1 and 2 for $P$-wave amplitudes $|L_u|^2$ and $|L_d|^2$ for $-t$=0.005-0.20 (GeV/c)$^2$.}
\label{Figure 1.}
\end{figure}

\begin{figure}[hp]
\includegraphics[width=12cm,height=10.5cm]{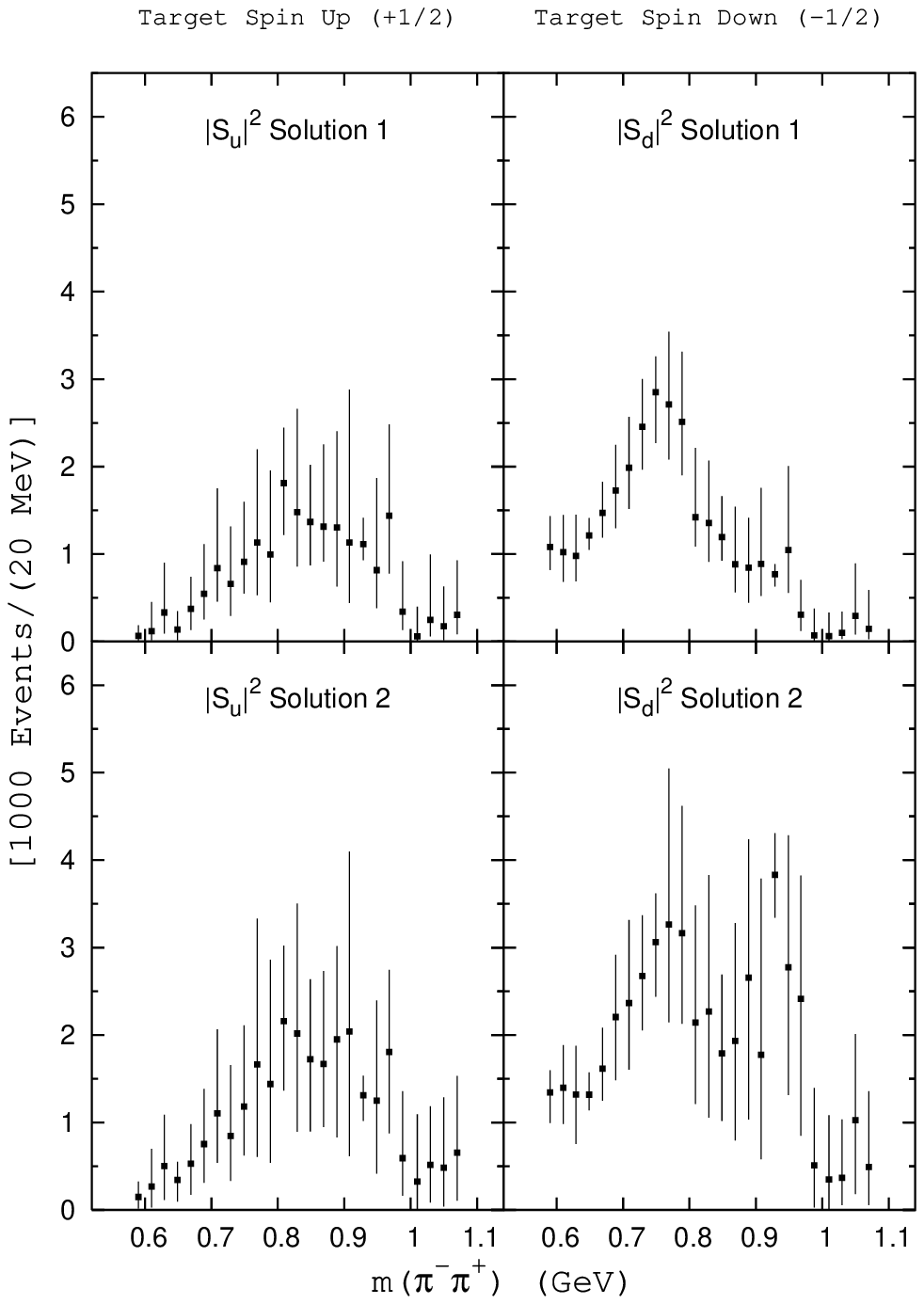}
\caption{Solutions 1 and 2 for $S$-wave amplitudes $|S_u|^2$ and $|S_d|^2$ for $-t$=0.005-0.20 (GeV/c)$^2$.}
\label{Figure 2.}
\end{figure}

\begin{figure}[htp]
\includegraphics[width=12cm,height=10.5cm]{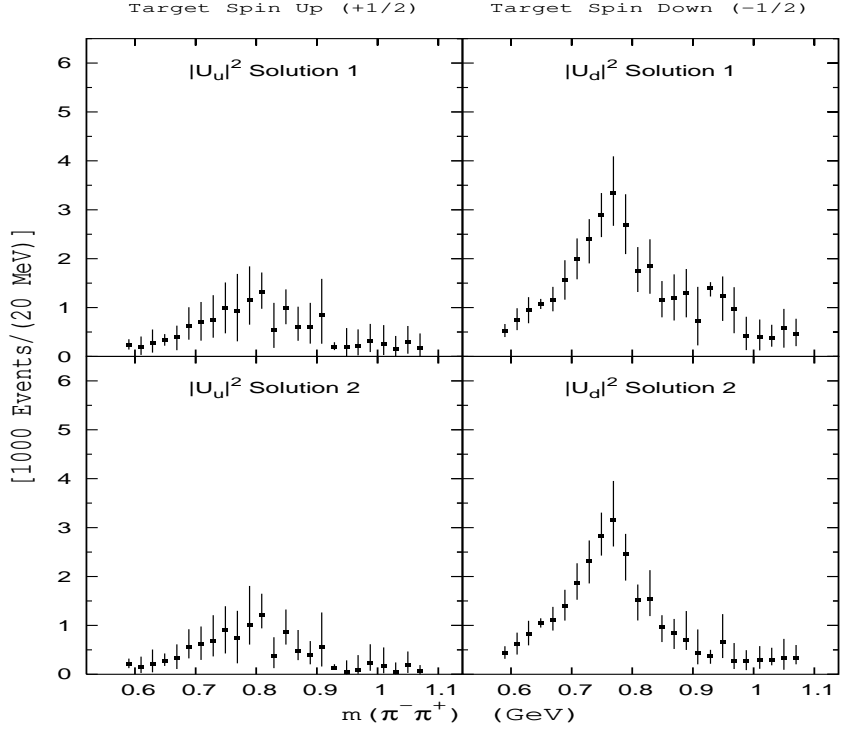}
\caption{Solutions 1 and 2 for $P$-wave amplitudes $|U_u|^2$ and $|U_d|^2$ for $-t$=0.005-0.20 (GeV/c)$^2$.}
\label{Figure 3.}
\end{figure}

\begin{figure}[h]
\includegraphics[width=12cm,height=10.5cm]{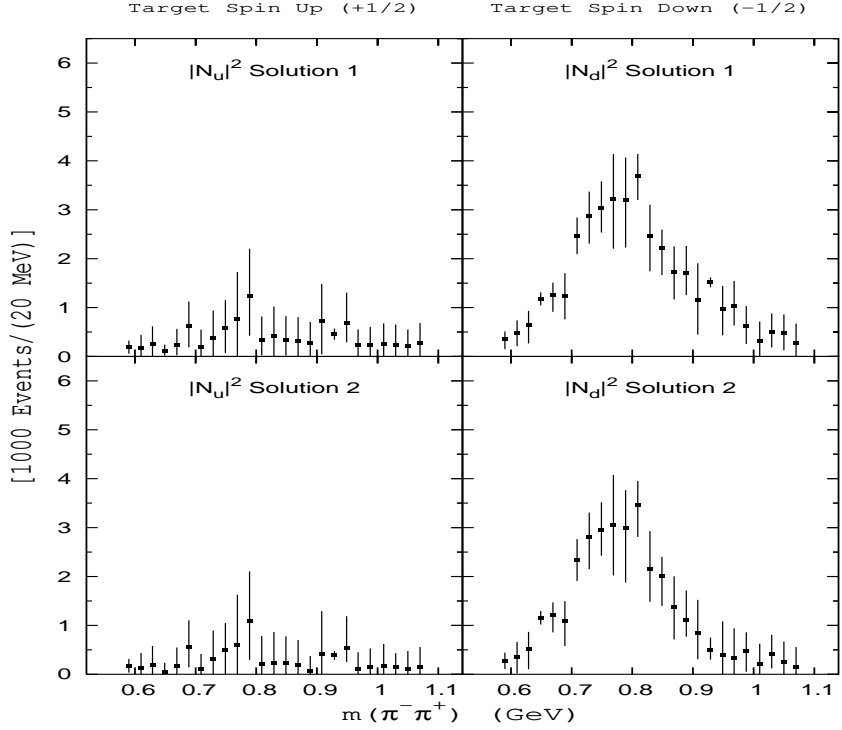}
\caption{Solutions 1 and 2 for $P$-wave amplitudes $|N_u|^2$ and $|N_d|^2$ for $-t$=0.005-0.20 (GeV/c)$^2$.}
\label{Figure 4.}
\end{figure}

For each set of solutions for moduli $|A(i)|,|\overline {A}(j)|,i,j=1,2$ we calculate from (2.11) and (2.12) the cosines
\begin{eqnarray} 
\cos(\Phi_{LS}(i))={a_4 \over {|L(i)||S(i)|}}\\
\nonumber 
\cos(\Phi_{US}(i))={a_5 \over {|U(i)||S(i)|}}\\
\nonumber
\cos(\Phi_{LU}(i))={a_6 \over {|L(i)||U(i)|}}\\
\nonumber
\end{eqnarray}
with similar equations for $\cos(\overline {\Phi}_{LS}(j))$, $\cos(\overline {\Phi}_{US}(j))$, $\cos(\overline {\Phi}_{LU}(j))$. From the phase condition
\begin{equation}
\Phi_L-\Phi_S=(\Phi_U-\Phi_S)+(\Phi_L-\Phi_U)
\end{equation}
we obtain 3 equations for sines
\begin{eqnarray}
C_A=\sin(\Phi_{LS}) \sin(\Phi_{US})=+\cos(\Phi_{LU})-\cos(\Phi_{LS}) \cos(\Phi_{US})\\
\nonumber
C_B=\sin(\Phi_{LS}) \sin(\Phi_{LU})=+\cos(\Phi_{US})-\cos(\Phi_{LS}) \cos(\Phi_{LU})\\
\nonumber
C_C=\sin(\Phi_{US}) \sin(\Phi_{LU})=-\cos(\Phi_{LS})+\cos(\Phi_{US}) \cos(\Phi_{LU})\\
\nonumber
\end{eqnarray}
The system is solvable provided that sign($C_C$)=sign($C_A C_B$). Given $\sin(\Phi_{LS})$,  the $\sin(\Phi_{US})$ and $\sin(\Phi_{LU})$ can be calculated from $C_A$ and $C_B$, respectively. With cosine and sines known, the phases can be calculated. With
\begin{equation}
\sin(\Phi_{LS})= \epsilon \sqrt{1-\cos^2(\Phi_{LS})}, \quad 
\sin(\overline{\Phi}_{LS})= \overline {\epsilon} \sqrt{1-\cos^2(\overline {\Phi}_{LS})}
\end{equation}
where the signs $\epsilon=\pm 1, \overline {\epsilon}=\pm 1$. The amplitude analyses assumed positive root for both sines. Such procedure is uniqe provided that the phases $\Phi_{LS}$ and $\overline {\Phi}_{LS}$ do not change signs. A change of sign would manifest itself as a double zero in the phases $\Phi_{LS}$ and $\overline {\Phi}_{LS}$ calculated from the positive roots for the sines. In all 8 analyzed reactions in Table IV. we found no evidence of a double zero in any of the phases $\Phi_{LS}$ and $\overline {\Phi}_{LS}$ so calculated and apart from the sign ambiguity these phases are uniquely determined.\\

The mass dependence at small $-t$ and the $t$-dependence at $\rho^0(770)$ mass region of $\Phi_{LS}$ and $\overline {\Phi}_{LS}$ in $\pi^- p \to \pi^- \pi^+ n$ are shown in Figures 5 and 6, respectively. Figures 7 and 8 show the mass dependence of the phases $\Phi_{US}, \overline {\Phi}_{US}$ and $\Phi_{LU}, \overline {\Phi}_{LU}$, repectively. As the figures show, these phases are continous and nearly constant functions. From this fact we can understand why the phases $\Phi_{LS}, \overline {\Phi}_{LS}$ cannot change sign. From (3.6) we see that a change of signs of $\Phi_{LS}$ or $ \overline {\Phi}_{LS}$ results in the change of sign of all phases of the same transversity. This would result in large unphysical discontinuities in the phases $\Phi_{US}, \overline {\Phi}_{US}$ or $\Phi_{LU}, \overline {\Phi}_{LU}$.\\

In the next Section IV. we show that the phases $\Phi_{LS}$ and $\overline {\Phi}_{LS}$ are closely related to the positivity of the reduced density matrices $\rho^0(P_y)=\rho^0_u+P_y \rho^0_y$ for pure initial states $P_y= \pm 1$. The fact that these phases do not change signs implies that the eigenvalues of the density matrices $\rho^0(P_y= \pm1)$ are all non-zero.\\

For negative roots in (3.7) all phases change signs. There are four combinations of the
signs $\epsilon$ and $\overline {\epsilon}$. The phases with opposite signs correspond to
complex conjugate reduced transversity amplitudes. For any given solution of moduli $|A(i)|,
|\overline {A}(j)|,i,j=1,2$ there is a four-fold ambiguity in the phases of the reduced transversity amplitudes $A=L,U$
\begin{equation}
L=|L|\exp (i\epsilon \Phi_{LS}), \quad \overline {L}=|\overline {L}|\exp (i\overline {\epsilon} \overline {\Phi}_{LS})
\end{equation}
\[
U=|U|\exp (i\epsilon \Phi_{US}), \quad \overline {U}=|\overline {U}|\exp (i\overline {\epsilon} \overline {\Phi}_{US})
\]
In the Section V. we show how the four-fold sign ambiguity can be resolved into a single set of phases for each set of moduli $|A(i)|,|\overline {A}(j)|,i,j=1,2$ in measurements of additional information on polarized targets.

\begin{figure}[t]
\includegraphics[width=12cm,height=10.5cm]{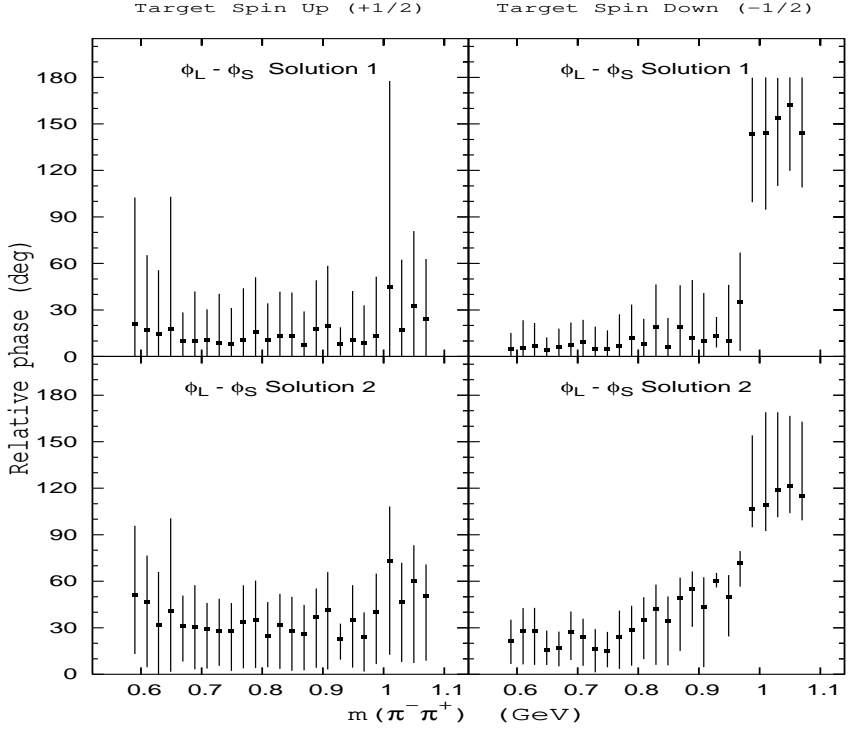}
\caption{Solutions 1 and 2 for relative phases $\Phi_{LS}=\Phi_{L_u}-\Phi_{S_u}$ and 
$\overline {\Phi}_{LS}=\Phi_{L_d}-\Phi_{S_d}$ for $-t$=0.005-0.20 (GeV/c)$^2$.}
\label{Figure 5.}
\end{figure}

\begin{figure}[h]
\includegraphics[width=12cm,height=10.5cm]{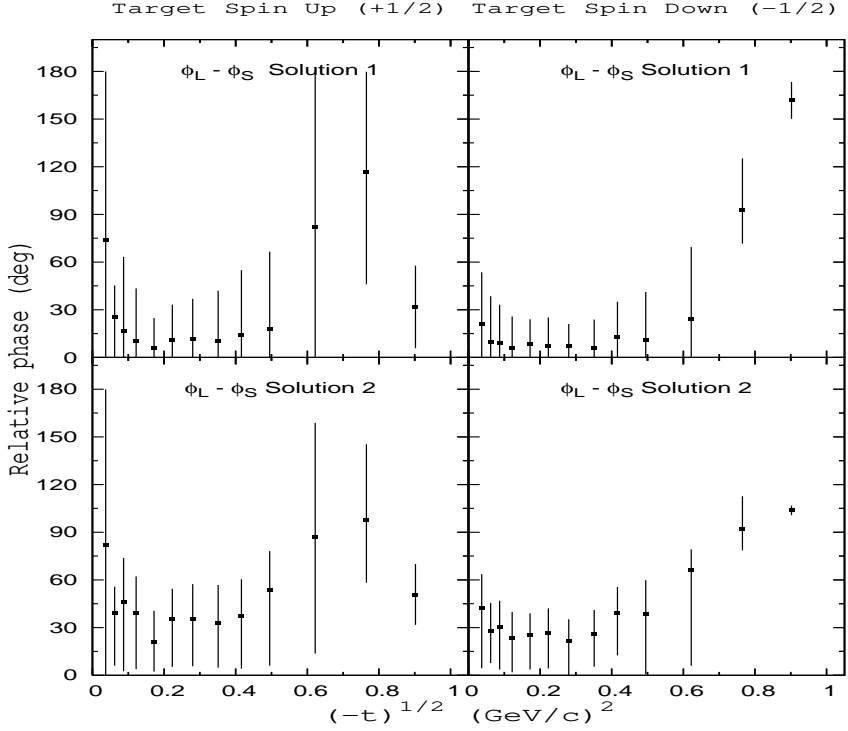}
\caption{Solutions 1 and 2 for relative phases $\Phi_{LS}=\Phi_{L_u}-\Phi_{S_u}$ and 
$\overline {\Phi}_{LS}=\Phi_{L_d}-\Phi_{S_d}$ in $\rho^0(770)$ mass region of $m$=710-830 MeV.}
\label{Figure 6.}
\end{figure}

\begin{figure}[t] 
\includegraphics[width=12cm,height=10.5cm]{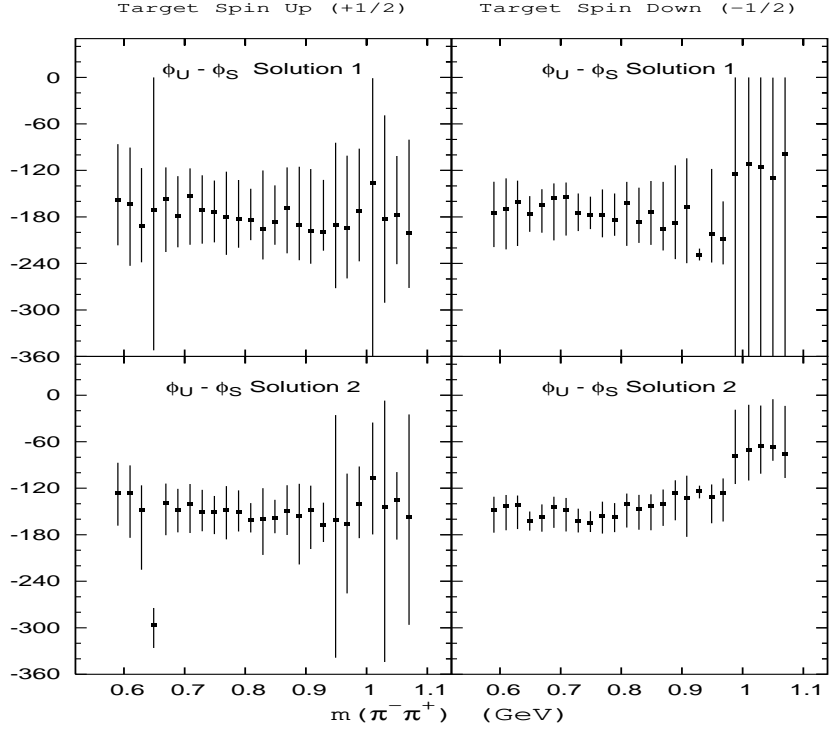}
\caption{Solutions 1 and 2 for relative phases $\Phi_{US}=\Phi_{U_u}-\Phi_{S_u}$ and 
$\overline {\Phi}_{US}=\Phi_{U_d}-\Phi_{S_d}$ for $-t$=0.005-0.20 (GeV/c)$^2$.}
\label{Figure 7.}
\end{figure}

\begin{figure}[h]
\includegraphics[width=12cm,height=10.5cm]{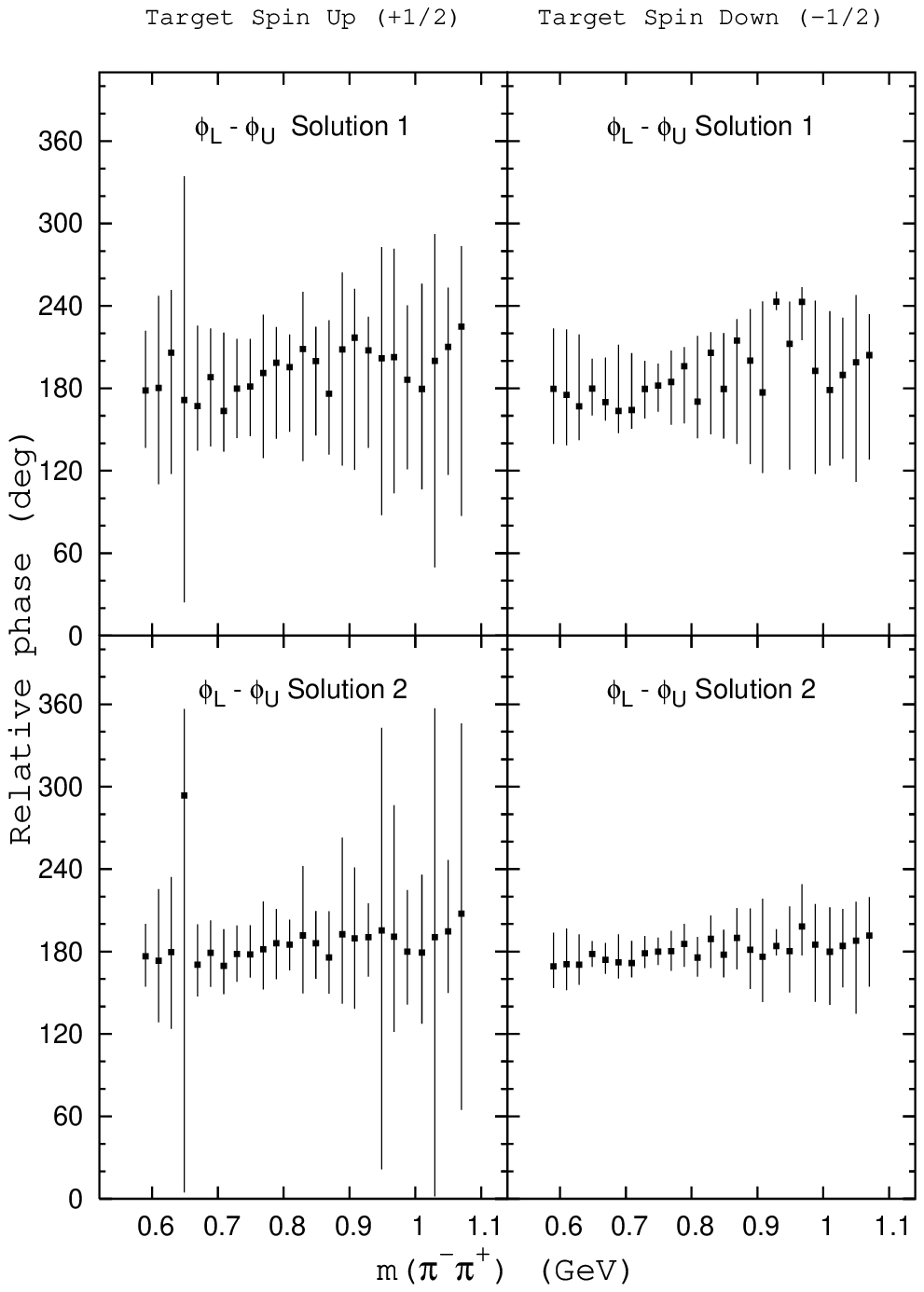}
\caption{Solutions 1 and 2 for relative phases $\Phi_{LU}=\Phi_{L_u}-\Phi_{U_u}$ and 
$\overline {\Phi}_{LU}=\Phi_{L_d}-\Phi_{U_d}$ for $-t$=0.005-0.20 (GeV/c)$^2$.}
\label{Figure 8.}
\end{figure}

\section{Positivity and decoherence constraints on reduced $S$- and $P$-wave density matrix and the phases $\Phi_{LS}$ and $\overline {\Phi}_{LS}$.}

The measured density matrix elements $Re (\rho^0_u)^{JJ'}_{\lambda \lambda'}$ and 
$Re (\rho^0_y)^{JJ'}_{\lambda \lambda'}$ are real parts of elements of complex density matrices $(\rho^0_u)^{JJ'}_{\lambda \lambda'}$ and $(\rho^0_y)^{JJ'}_{\lambda \lambda'}$. 
The matrices $(\rho^0_u)^{JJ'}_{\lambda \lambda'}$ and $(\rho^0_y)^{JJ'}_{\lambda \lambda'}$
are hermitian and so is their measured combination for target polarization component $P_y$
\begin{equation}
\rho^0(P_y)^{JJ'}_{\lambda \lambda'}=
(\rho^0_u)^{JJ'}_{\lambda \lambda'}+P_y(\rho^0_y)^{JJ'}_{\lambda \lambda'}
\end{equation}
We will be interested in two specific cases of pure initial states $P_y=P_\tau= \pm 1$. Omitting the labels ${}^0_u$, ${}^0_y$, the argument $(P_y)$ and suppressing the indices $JJ'$, these reduced matrices have the following general form for the $S$- and $P$-wave subspace
\begin{equation}
\mathbf {\rho^0} =
\left( \begin{array}{cccc}
\rho_{ss} & \rho_{s1} & \rho_{s0} & \rho_{s-1}\\
\rho_{1s} & \rho_{11} & \rho_{10} & \rho_{1-1}\\
\rho_{0s} & \rho_{01} & \rho_{00} & \rho_{0-1}\\
\rho_{-1s} & \rho_{-11} & \rho_{-10} & \rho_{-1-1}\\
\end{array} \right)
\end{equation}
Using relations due to hermiticity and $P$-parity conservation~\cite{svec07a} 
\begin{eqnarray}
(\rho^0_k)^{J'J}_{\lambda' \lambda} & = & ((\rho^0_k)^{JJ'}_{\lambda \lambda'})^*\\
\nonumber
(\rho^0_k)^{JJ'}_{-\lambda -\lambda'} & = & 
(-1)^{\lambda +\lambda'}(\rho^0_k)^{JJ'}_{\lambda \lambda'}
\nonumber
\end{eqnarray}
for $k=u,y$ and similar relations for $(\rho^0(P_y))^{JJ'}_{\lambda \lambda'}$, the matrices 
$\rho^0$ take the form
\begin{equation}
\mathbf {\rho^0} =
\left( \begin{array}{cccc}
\rho_{ss} & \rho_{1s}^* & \rho_{0s}^* & -\rho_{1s}^*\\
\rho_{1s} & \rho_{11} & \rho_{01^*} & \rho_{1-1}\\
\rho_{0s} & \rho_{01} & \rho_{00} & -\rho_{01}\\
-\rho_{1s} & \rho_{1-1} & -\rho_{01}^* & \rho_{11}\\
\end{array} \right)
\end{equation}
The positivity of the hermitiam matrices $\rho^0$ requires that 
$\det {\rho^0} \geq 0$~\cite{bourrely80,leader01}. Calculations show that
\begin{equation}
\det {\rho^0}=
(\rho_{11}+\rho_{1-1}) \Bigl ( \rho_{ss} \rho_{00}(\rho_{11}-\rho_{1-1}) 
-2 \rho_{ss}|\rho_{01}|^2-
\end{equation}
\[
(\rho_{11}-\rho_{1-1})|\rho_{0s}|^2-2 \rho_{00} |\rho_{1s}|^2
+4|\rho_{0s}||\rho_{1s}||\rho_{01}| \Bigr )
\]
We split the determinant into two parts $\det{\rho^0}=D+P$ where
\begin{equation}
D=(\rho_{11}+\rho_{1-1}) \Bigl ( \rho_{ss} \rho_{00}(\rho_{11}-\rho_{1-1}) 
-2 \rho_{ss}(Re \rho_{01})^2-
\end{equation}
\[
(\rho_{11}-\rho_{1-1})(Re \rho_{0s})^2-2 \rho_{00} (Re \rho_{1s})^2
+4 Re \rho_{0s} Re\rho_{1s}Re\rho_{01} \Bigr )
\]
\begin{equation}
P=(\rho_{11}+\rho_{1-1}) \Bigl ( 4|\rho_{0s}||\rho_{1s}||\rho_{01}|
-2 \rho_{ss}(Im \rho_{01})^2-
\end{equation}
\[
(\rho_{11}-\rho_{1-1})(Im \rho_{0s})^2-2 \rho_{00} (Im \rho_{1s})^2
-4Re \rho_{0s} Re \rho_{1s} Re \rho_{01} \Bigr )
\]
It is of some interest to note that in general
\begin{equation}
\det{Re \rho^0}=D, \quad \det{Im \rho^0}=0
\end{equation}

Next we focus on reduced density matrices $\rho^0(P_y)=\rho^0_u+P_y \rho^0_y$ with pure initial states 
$P_y= \pm1$. For $P_y=+1$ we get from the Table I.
\begin{equation}
\rho_{ss}=|S|^2, \quad \rho_{00}=|L|^2, \quad \rho_{11}-\rho_{1-1}=|U|^2, \quad 
\rho_{11}+\rho_{1-1}=|N|^2
\end{equation}
The expression in the large parenthesis in the term $D(P_y=+1)$ is identical to the phase condition (2.16) so that $D(P_y=+1)=0$ as the result of the assumption that the $S$- and $P$-wave subsystem forms a decoherence free subspace of the reduced density matrix $\rho^0_f$. Using the relations
\begin{eqnarray}
Im \rho_{0s} & = & |L||S| \sin \Phi_{LS}\\
\nonumber
\sqrt{2} Im \rho_{1s} & = & |U||S| \sin \Phi_{US}\\
\nonumber
\sqrt{2} Im \rho_{01} & = & |L||U| \sin \Phi_{LU}
\nonumber
\end{eqnarray}
and the phase condition (2.15) for the cosines to simplify the term $P(P_y=+!)$, the determinant for $P_y= + 1$ takes the form
\begin{equation}
\det{\rho^0(+1)}=|N|^2|U|^2|L|^2|S|^2 (\sin \Phi_{LS})^2
\end{equation}
We obtain a similar expression for the determinant for $P_y=-1$ 
\begin{equation}
\det{\rho^0(-1)}=|\overline {N}|^2|\overline {U}|^2|\overline {L}|^2|\overline {S}|^2 (\sin \overline {\Phi}_{LS})^2
\end{equation}
The determinants are nonnegative as follows from the positivity of the density matrices $\rho^0(P_y)$.  The density matrices are hermitian matrices and thus can be diagonalized. If all their eigenvalues are non-zero, the determinants will be strictly positive. This would imply that the phases $\Phi_{LS}$ and $\overline {\Phi}_{LS}$ cannot change sign, as was actually observed in Section III. from the analysis of the experimental data.\\

\section{Resolution of sign ambiguities in the phases of reduced $S$- and $P$-wave transversity amplitudes in measurements of $Im \rho^0_x$ and $Im \rho^0_z$.} 

The $S$- and $P$-wave density matrix elements $Im \rho^0_x$ and $Im \rho^0_z$ involve interferences between reduced unnatural and natural exchange amplitudes of opposite transversity shown in Table II.. First we shall show that the elements 
$Im (\rho^0_x)^{01}_{s1}$ and $Im (\rho^0_z)^{01}_{s1}$ determine the phases $\alpha_N=\Phi_{N_u}-\Phi_{S_d}$ and $\overline{\alpha}_N=\Phi_{N_d}-\Phi_{S_u}$ of the amplitudes $N$ and $\overline {N}$ with a two-fold ambiguity for each set of moduli $|N(i)|,|\overline {N}(j)|,i,j,=1,2$. From the Table II. and (2.5) we have
\begin{equation}
r_1=\sqrt{2}Im (\rho^0_x)^{01}_{s1} \Sigma = Re(-|S|\overline {N}^*+N|\overline{S}|)=
-|S||\overline {N}| \cos \overline {\alpha}_N+|N||\overline {S}| \cos \alpha_N
\end{equation}
\[ 
s_1=\sqrt{2}Im (\rho^0_z)^{01}_{s1} \Sigma = Im(+|S|\overline {N}^*-N|\overline{S}|)=
-|S||\overline {N}| \sin \overline {\alpha}_N-|N||\overline {S}| \sin \alpha_N
\]
where $\Sigma = d^2 \sigma /dmdt$. From (5.1) we can solve for $\cos \alpha_N$ and $\sin \alpha_N$
\begin{eqnarray}
\cos \alpha_N & = & {1 \over{|N||\overline{S}|}} \Bigl (r_1+|S||\overline {N}| \cos \overline {\alpha}_N \Bigr ) \\
\nonumber 
\sin \alpha_N & = & {-1 \over{|N||\overline{S}|}} \Bigl (s_1+|S||\overline {N}| \sin \overline {\alpha}_N \Bigr ) 
\nonumber 
\end{eqnarray}
Substituting into $\cos^2 \alpha_N+\sin^2 \alpha_N=1$ we find
\begin{equation}
\cos \overline {\alpha}_N={1\over{2|S||\overline {N}|r_1}} \Bigl ( A-2|S||\overline{N}|s_1 \sin \overline {\alpha}_N \Bigr )
\end{equation}
where
\[
A=|N|^2|\overline {S}|^2 -|S|^2|\overline {N}|^2-(r_1^2+s_1^2)
\]
Substituting into $\cos^2 \overline {\alpha}_N+\sin^2 \overline {\alpha}_N=1$ yields a quadratic equation for $\sin \overline {\alpha}_N$ with two solutions
\begin{equation}
\sin \overline {\alpha}_N = {A \over {2|S||\overline {N}|(r_1^2+s_1^2)}} 
\Bigl ( s_1 \pm r_1 \sqrt{B-1} \Bigr )
\end{equation}
where
\[
B={{4|S|^2| \overline{N}|^2(r_1^2+s_1^2)} \over {A^2}}
\]
From (5.4) we can now calculate $\cos \overline {\alpha}_N$
\begin{equation}
\cos \overline {\alpha}_N= {A \over {2|S||\overline {N}|(r_1^2+s_1^2)}} 
\Bigl ( r_1 \mp s_1 \sqrt{B-1} \Bigr ) 
\end{equation}
and from (5.3) we obtain
\begin{eqnarray}
\cos \alpha_N & = &  {A \over {2|N||\overline {S}|(r_1^2+s_1^2)}} 
\Bigl ( r_1C \mp s_1 \sqrt{B-1} \Bigr )\\
\nonumber
\sin \alpha_N & = &  {{-A} \over {2|N||\overline {S}|(r_1^2+s_1^2)}} 
\Bigl ( s_1C \pm r_1 \sqrt{B-1} \Bigr )
\nonumber
\end{eqnarray}
where
\[
C={{A+2(r_1^2+s_1^2)} \over {A}}
\]
Note that for each set of moduli $|A(i)|,|\overline{A}(j)|$, $i,j=1,2$ the two solutions for the phases $\alpha_N(ij), \overline {\alpha}_N(ij)$ will depend on the set.\\

We shall now show that only one solution is consistent with the remaining equations
from the Table II. for each set of the moduli and that this solution resolves the four-fold sign ambiguity. To this end we introduce a convenient notation for real and imaginary parts of reduced transversity amplitudes $A=L,U,N$ 
\begin{eqnarray}
A=A_1+iA_2 & = & |A| \cos \alpha_A + i|A| \sin \alpha_A\\
\nonumber
\overline {A} =\overline {A}_1+i\overline {A}_2 & = & 
|\overline {A}| \cos \overline {\alpha}_A+i|\overline {A}| \sin \overline {\alpha}_A
\nonumber
\end{eqnarray}
where the phases $\alpha_A$ and $\overline {\alpha}_A$ for amplitudes $A=L,U$ are given by (3.8). The equations for the remaining density matrix elements from the Table II. then take the form
\begin{eqnarray}
r_2= \sqrt{2} Im ( \rho^0_x)^{11}_{01} \Sigma & = & 
\overline {L}_1 N_1+ \overline {L}_2 N_2-L_1 \overline {N}_1 -L_2 \overline {N}_2 \\
\nonumber
s_2= \sqrt{2} Im ( \rho^0_z)^{11}_{01} \Sigma & = & 
\overline {L}_2 N_1- \overline {L}_1 N_2+L_2 \overline {N}_1 -L_1 \overline {N}_2 
\nonumber
\end{eqnarray}
\begin{eqnarray}
r_3=-Im ( \rho^0_x)^{11}_{-11} \Sigma & = & 
\overline {U}_1 N_1+ \overline {U}_2 N_2-U_1 \overline {N}_1 -U_2 \overline {N}_2 \\
\nonumber
s_3=-Im ( \rho^0_z)^{11}_{-11} \Sigma & = & 
\overline {U}_2 N_1- \overline {U}_1 N_2+U_2 \overline {N}_1 -U_1 \overline {N}_2 
\nonumber
\end{eqnarray}

Suppose that for a given set of moduli $|A(i)|,|\overline {A}(j)|$ the equations (5.8) and (5.9) are satisfied by one of the two solutions for the phases $\alpha_N(ij), \overline {\alpha}_N(ij)$ for a particular set of signs of phases $\alpha_A(i), \overline {\alpha}_A(j), A=L,U$ from (3.8). To be specific, let us suppose this solution for $N, \overline {N}$ is the solution with the positive sign in (5.4) and let us label all amplitudes in this solution $A^+,\overline {A}^+$. We now show that the amplitudes $N^-,\overline {N}^-$ for the other solution with the negative sign in (5.4) cannot be a solution of (5.8) and (5.9) for any choice of signs of phases $\alpha_A(i), \overline {\alpha}_A(j), A=L,U$ in (3.8).\\

To prove this statetement let us suppose the contrary and assume that the amplitudes $N^-,\overline {N}^-$ satisfy (5.8) and (5.9) for some amplitudes $A^-,\overline{A}^-$, $A=L,U$ from (3.8). Since the amplitudes $A^-,\overline{A}^-$, $A=L,U$ differ from the amplitudes $A^+,\overline{A}^+$, $A=L,U$ only in the sign of phases, their real parts are the same and the imaginary parts differ at most by sign
\begin{equation}
A_2^-=\lambda A_2^+, \quad \overline {A}_2^-=\overline{ \lambda} \overline {A}_2^+, \quad A=L,U
\end{equation}
where $\lambda= \pm 1, \overline {\lambda}= \pm 1$. In the next step we subtract the set of equations with amplitudes $N^-,\overline {N}^-$ from the set with amplitudes $N^+,\overline {N}^+$ and get a homogeneous set of equations
\begin{eqnarray}
\overline {L}_1^+(N_1^+-N_1^-)+\overline{L}_2^+(N_2^+-\overline{\lambda}N_2^-)-
L_1^+(\overline {N}_1^+-\overline {N}_1^-)-L_2^+(\overline{N}_2^+-\lambda \overline {N}_2^-) & = & 0\\
\nonumber
\overline {L}_2^+(N_1^+-\overline{\lambda}N_1^-)-\overline{L}_1^+(N_2^+-N_2^-)+
L_2^+(\overline {N}_1^+-\lambda \overline {N}_1^-)-L_1^+(\overline{N}_2^+-\overline {N}_2^-) & = & 0
\nonumber
\end{eqnarray}
\begin{eqnarray}
\overline {U}_1^+(N_1^+-N_1^-)+\overline{U}_2^+(N_2^+-\overline{\lambda}N_2^-)-
U_1^+(\overline {N}_1^+-\overline {N}_1^-)-U_2^+(\overline{N}_2^+-\lambda \overline {N}_2^-) & = & 0\\
\nonumber
\overline {U}_2^+(N_1^+-\overline{\lambda}N_1^-)-\overline{U}_1^+(N_2^+-N_2^-)+
U_2^+(\overline {N}_1^+-\lambda \overline {N}_1^-)-U_1^+(\overline{N}_2^+-\overline {N}_2^-) & = & 0
\nonumber
\end{eqnarray}
We now write (5.6) in the form
\begin{eqnarray}
\cos \alpha^\pm_N & = & {1\over {|N|}} \bigl ( a_1r_1 \mp a_2s_1 \bigr )\\
\nonumber
\sin \alpha^\pm_N & = & {1\over {|N|}} \bigl ( a_1s_1 \pm a_2r_1 \bigr )
\nonumber
\end{eqnarray}
and calculate differences and sums
\begin{equation}
N_1^+-N_1^- = -2a_2s_1, \quad N_1^++N_1^- = +2a_1r_1
\end{equation}
\[
N_2^+-N_2^- = +2a_2r_1, \quad N_1^++N_1^- = +2a_1s_1
\]
We can write similar expressions for amplitudes $\overline{N}^+$ and $\overline {N}^-$ by replacing the paramerters $a_1,a_2$ in (4.14) with parameters $\overline {a}_1, \overline {a}_2$ corresponding to the form (4.13) of the equations (4.5) and (4.4) for
 $\cos \overline {\alpha}_N$ and $\sin \overline {\alpha}_N$.\\

We now examine the equations (5.11) and (5.12) for each possible choice of $\lambda$ and $\overline {\lambda}$. The case $\lambda=\overline {\lambda}=+1$ is excluded as the 
amplitudes $N^-,\overline {N}^-$ cannot satisfy the same system as the amplitudes  
$N^+,\overline {N}^+$. For the case with $\lambda=\overline {\lambda}=-1$ the equations (5.11) and (5.12) take the form
\begin{eqnarray}
s_1 \Bigl (-\overline {L}_1^+a_2+\overline {L}_2^+a_1 +
L_1^+ \overline{a}_2 -L_2^+\overline {a}_1 \Bigr ) & = & 0\\
\nonumber
r_1 \Bigl (-\overline {L}_1^+a_2+\overline {L}_2^+a_1 -
L_1^+ \overline{a}_2 +L_2^+\overline {a}_1 \Bigr ) & = & 0\\
s_1 \Bigl (-\overline {U}_1^+a_2+\overline {U}_2^+a_1 +
U_1^+ \overline{a}_2 -U_2^+\overline {a}_1 \Bigr ) & = & 0\\
\nonumber
r_1 \Bigl (-\overline {U}_1^+a_2+\overline {U}_2^+a_1 -
U_1^+ \overline{a}_2 +U_2^+\overline {a}_1 \Bigr ) & = & 0
\nonumber
\end{eqnarray}
The terms in the parentheses are all different. In particular, the large differences between the moduli $|L|,|\overline {L}|$ and $|U|,|\overline {U}|$ mean large differences in the parentheses for $r_1$ at least one of which must be non-zero. Since $r_1$ has been measured in the CERN experiments on polarized targets and is non-zero, the equations (5.15) and (5.16) cannot be satisfied and this case is excluded.\\

For the case $\lambda=+1$ and $\overline {\lambda}=-1$ the equations  (5.11) and (5.12) read
\begin{eqnarray}
\Bigl (-\overline {L}_1^+a_2+\overline {L}_2^+a_1 +
L_1^+ \overline{a}_2 \Bigr )s_1 -\Bigl (L_2^+\overline {a}_2 \Bigr )r_1 & = & 0\\
\nonumber
\Bigl (-\overline {L}_1^+a_2+\overline {L}_2^+a_1 -
L_1^+ \overline{a}_2 \Bigr )r_1 -\Bigl (L_2^+\overline {a}_2 \Bigr )s_1 & = & 0\\
\Bigl (-\overline {U}_1^+a_2+\overline {U}_2^+a_1 +
U_1^+ \overline{a}_2 \Bigr )s_1 -\Bigl (U_2^+\overline {a}_2 \Bigr )r_1 & = & 0\\
\nonumber
\Bigl (-\overline {U}_1^+a_2+\overline {U}_2^+a_1 -
U_1^+ \overline{a}_2 \Bigr )r_1 -\Bigl (U_2^+\overline {a}_2 \Bigr )s_1 & = & 0
\nonumber
\end{eqnarray}
Combining the first two and the last two equations we obtain two equations of interest
\begin{eqnarray}
\Bigl (-\overline {L}_1^+a_2+\overline {L}_2^+a_1 +L_1^+ \overline{a}_2 \Bigr )s_1^2 & = & 
\Bigl (-\overline {L}_1^+a_2+\overline {L}_2^+a_1 -L_1^+ \overline{a}_2 \Bigr )r_1^2\\
\Bigl (-\overline {U}_1^+a_2+\overline {U}_2^+a_1 +U_1^+ \overline{a}_2 \Bigr )s_1^2 & = & 
\Bigl (-\overline {U}_1^+a_2+\overline {U}_2^+a_1 -U_1^+ \overline{a}_2 \Bigr )r_1^2
\nonumber
\end{eqnarray}
The terms in the parentheses are all different. The parentheses for amplitudes $L, \overline {L}$ and amplitudes $U, \overline {U}$ are not proportional to each other since the phases of $L, \overline {L}$ and $U, \overline {U}$ are 180$^o$ out of phase, as seen in Fig. 8. Moreover, as we shall discuss in Section VII., the moduli of these amplitudes have different $\rho^0(770)$ widths and structures around $f_0(980)$. The equations (5.17) and (5.18) thus cannot be satisfied and this case is excluded. The analysis of the case $\lambda=-1$ and $\overline{\lambda}=+1$ is similar with the same conclusion. The solution $N^+, \overline {N}^+$ selects a unique set of phases of amplitudes $L,U$ and $\overline{L}, \overline {U}$. The equations (5.8) and (5.9) change when the phases of these amplitudes change sign. The change results in a different solution for amplitudes $N, \overline {N}$ which is not compatible with the data in (5.1).\\

We conclude that the measurements of $S$-and $P$-wave density matrix elements $Im \rho^0_x$ and $Im \rho^0_z$ listed in Table II. unambiguously select a unique solution for $S$- and $P$-wave reduced transversity amplitudes. In effect, for any set $i,j$ of solutions for the moduli there exists only one set of signs of phases in (3.8) such that the solution of equations (5.8) and (5.9) for $N_1,N_2,\overline{N}_1, \overline{N}_2$ satisfies the conditions $|N(i)|^2=N_1^2+N_2^2$ and $|N(j)|^2=\overline{N}_!^2+\overline{N}_2^2$. These measurements were not feasible in the 1970's when the first CERN measurements on polarized targets were done. With the advent of advanced frozen spin targets the direction of target polarization can be selected at will with high degree of polarization~\cite{leader01}. The next generation of measurements of pion creation processes on polarized targets will therefore provide data on the complete reduced density matrix.

\section{Evidence for $\rho^0(770)-f_0(980)$ mixing in $\pi^- p \to \pi^- \pi^+ n$ and $\pi^+ n \to \pi^+ \pi^- p$.}

We first recall the equations (2.17) for the moduli $|S|^2$, $|U|^2$ and $|N|^2$ 
\begin{eqnarray}
|S|^2 = (a_1+a_2)-3|L|^2\\
\nonumber
|U|^2=|L|^2-{1 \over {2}}(a_2+a_3)\\
\nonumber
|N|^2=|L|^2-{1 \over {2}}(a_2-a_3)
\end{eqnarray}
The data components $a_1+a_2$, ${1 \over {2}}(a_2+a_3)$ and  ${1 \over {2}}(a_2-a_3)$ were calculated from those Monte Carlo samplings of density matrix elements in the error volume of the data for which physical solutions for the amplitudes were found. The results for $a_1+a_2$ and  ${1 \over {2}}(a_2+a_3)$ are shown in Figure 9. The results for ${1 \over {2}}(a_2-a_3)$ are very similar to those of ${1 \over {2}}(a_2+a_3)$ and are not shown. The Figure 9 shows pronounced $\rho^0(770)$ peaks for target spin "down" components and a suppression of $\rho^0(770)$ in the target spin "up" components. There is a pronounced dip at 970 MeV in target spin "down" components and a dip at 1030 MeV in target spin "up" components corresponding to the $f_0(980)$ resonance.\\

\begin{figure}[h]
\includegraphics[width=12cm,height=10.5cm]{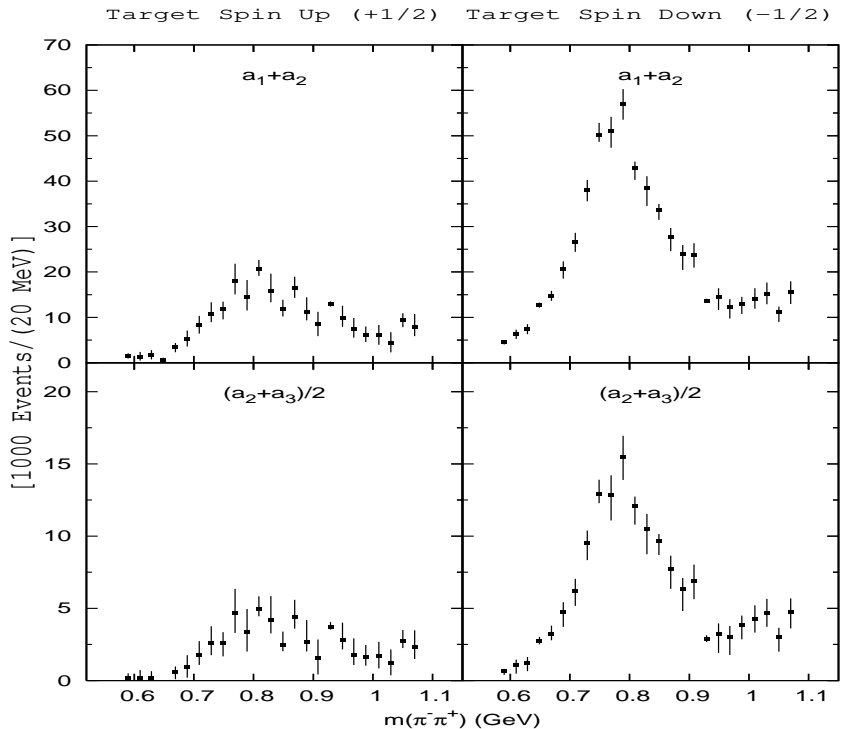}
\caption{Data components $a_1+a_2$ of the moduli $|S_\tau|^2$ and $(a_2+a_3)/2$ for the moduli $|U_\tau|^2$, $\tau=u,d$ for $-t$=0.005-0.20 (GeV/c)$^2$.}
\label{Figure 9.}
\end{figure}

Figure 1 shows the two solutions for the $P$-wave amplitudes $|L_u|^2$ and $|L_d|^2$. Remarkably, the two solutions of the two cubic equations reproduce closely the general features of the data components in the Figure 9. Figure 2 shows the two solutions for the $S$-wave amplitudes $|S_u|^2$ and $|S_d|^2$. The rho-like resonance and $f_0(980)$ are clearly resolved in both solutions for $|S_d|^2$, in contrast with our previous low resolution amplitude analyses based on 40,000 Monte Carlo 
samplings~\cite{svec96,svec97a,svec02a}. This peak at 770 MeV with a width at half-height of 155 MeV has its origin in the $\rho^0(770)$ peak of the data component $a_1+a_2$ which survives the subtraction of $3|L_d|^2$. Just like the $\rho^0(770)$ peak in data components 
${1 \over {2}}(a_2+a_3)$ and  ${1 \over {2}}(a_2-a_3)$ for target spin "down" contributes to $\rho^0(770)$ peaks in the amplitudes $|U_d|^2$ and $|N_d|^2$ (see Figures 3 and 4), so the $\rho^0(770)$ peak in the data component $a_1+a_2$ gives rise to the presence of $\rho^0(770)$ in the amplitudes $|S_d|^2$. Similarly we can see that the suppressed $\rho^0(770)$ peak in the data components for target spin "up" propagates to all target spin "up" moduli $|A_u|^2$, including $S$-wave amplitude $|S_u|^2$.\\

The data components $a_4$, $a_5$ and $a_6$ provide independent evidence for the presence of $\rho^0(770)$ in the $S$-wave amplitudes. Figures 5 shows relative phases $\Phi_{LS}= \Phi_{L_u}-\Phi_{S_u}$, $\overline {\Phi}_{LS}= \Phi_{L_d}-\Phi_{S_d}$. For target spin "up" the phase $\Phi_{LS}$ is constant and small, reflecting the similarity of $|S_u|^2$ and $|L_u|^2$. For target spin "down" the phase $\overline {\Phi}_{LS}$ is near zero in the Solution 1 and small and slowly varying in Solution 2 below 1000 MeV. These results indicate that the amplitudes $S_\tau$ and $L_\tau$, $\tau=u,d$ are in phase. Since the amplitudes $L_\tau$ resonate at $\rho^0(770)$, so must the amplitudes $S_\tau$.\\

This conclusion is in agreement with the phases $\Phi_{US}= \Phi_{U_u}-\Phi_{S_u}$, $\overline {\Phi}_{US}= \Phi_{U_d}-\Phi_{S_d}$ shown in Figure 7. These phases are nearly constant below 1000 MeV and indicate that the amplitudes $S_\tau$ and $U_\tau$ are 180$^o$ out of phase. Again, since $U_\tau$ resonate at $\rho^0(770)$, so must the amplitudes $S_\tau$. The Figure 8 shows that the relative phases of the amplitudes $L_\tau$ and $U_\tau$ are nearly constant and 180$^o$ out of phase, as expected from the two resonating amplitudes.\\

The observation of $\rho^0(770)-f_0(980)$ mixing in the $S$-wave amplitudes brings up the question of $\rho^0(770)-f_0(980)$ mixing in the $P$-wave amplitudes. Figure 1 shows a clear dip at 970 MeV in the amplitude $|L_d|^2$ followed by a sharp rise. The dip occurs at the mass of scalar resonance $f_0(980)$ and is more pronounced in the Solution 2. Figure 3 shows unexpected resonant structure (a bump) at the $f_0(980)$ mass in the Solution 1 of the amplitude $|U_d|^2$. These findings are a strong indication for $\rho^0(770)-f_0(980)$ mixing in the $P$-wave amplitudes. Since $f_0(980)$ resonance is produced by unnatural exchange, we expect $f_0(980)$ to be suppressed in natural exchange amplitudes $|N_\tau|^2$ as evidenced by the data in Figure 4.\\

Independent evidence for $\rho^0(770)-f_0(980)$ mixing in the $S$-wave amplitudes comes from the amplitude analysis of the CERN data on the $t$-dependence of the density matrix elements in the $\rho^0(770)$ mass region (reaction 2 in the Table IV.) for $-t$=0.0-1.0 GeV/c$^2$. In Figure 10 we compare the results for the amplitudes $|S_\tau|^2$ and $|L_\tau|^2$.  We find that the moduli have similar shapes in all solutions for both target spin "up" and "down" amplitudes. This will happen if the two amplitudes both resonate at $\rho^0(770)$.\\

\begin{figure}[h]
\includegraphics[width=12cm,height=10.5cm]{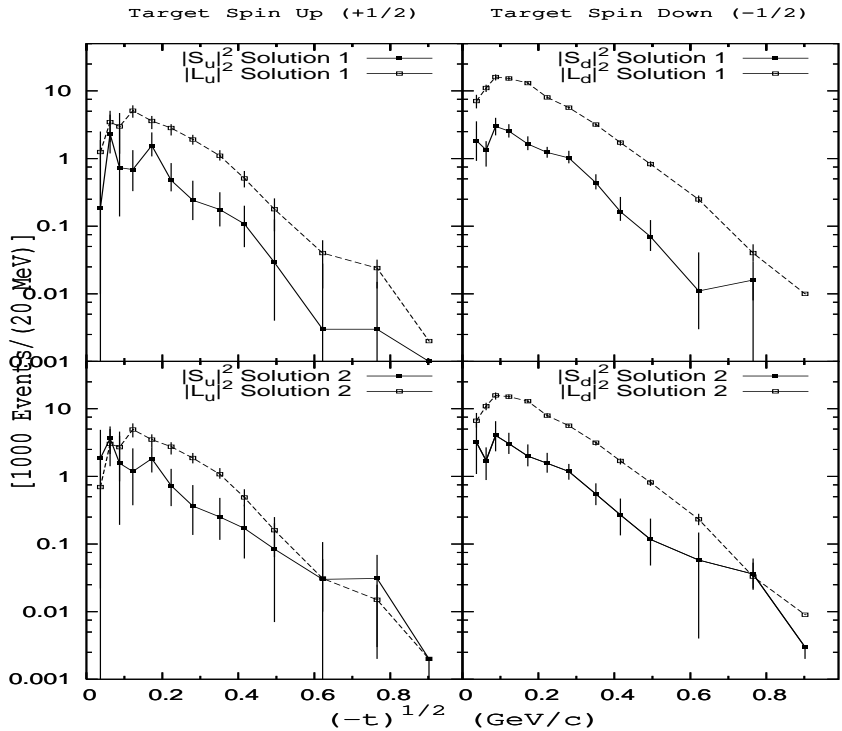}
\caption{Solutions 1 and 2 for the moduli $|S_\tau|^2$ (solid lines) and $|L_\tau|^2$ (dashed lines), $\tau=u,d$ in the $\rho^0(770)$ mass region $m$=710-830 MeV.}
\label{Figure 10.}
\end{figure}

\begin{figure}[h]
\includegraphics[width=12cm,height=10.5cm]{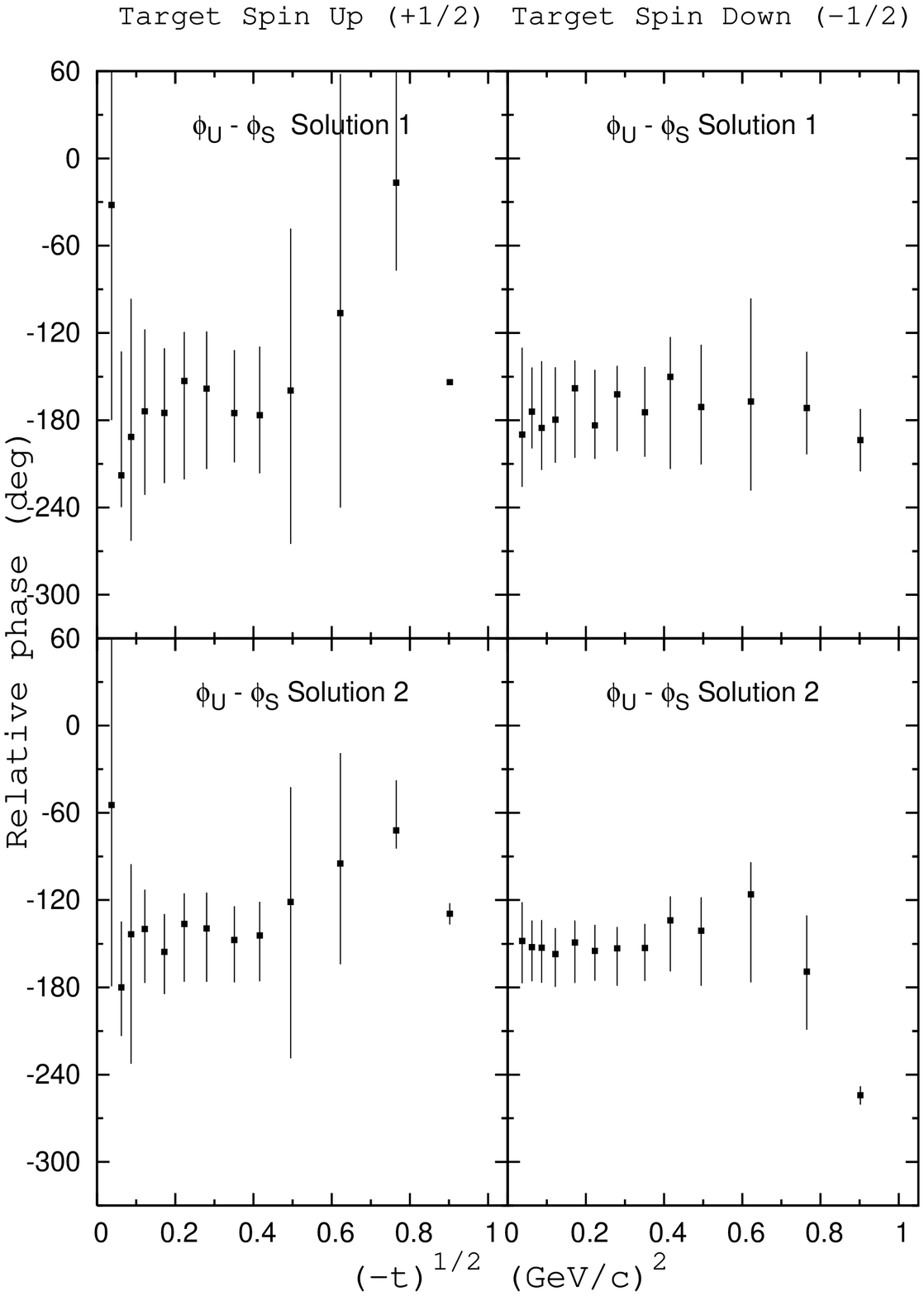}
\caption{Solutions 1 and 2 for relative phases $\Phi_{US}=\Phi_{U_u}-\Phi_{S_u}$ and 
$\overline {\Phi}_{US}=\Phi_{U_d}-\Phi_{S_d}$ in the $\rho^0(770)$ mass region $m$=710-830 MeV.}
\label{Figure 11.}
\end{figure}

This conclusion is supported by the results for the relative phases between amplitudes $S_\tau$ and $L_\tau$ in Figure 6 and amplitudes $S_\tau$ and $U_\tau$ in Figure 11. The phases $\Phi_{LS}= \Phi_{L_u}-\Phi_{S_u}$ and $\overline {\Phi}_{LS}= \Phi_{L_d}-\Phi_{S_d}$
are small and nearly constant below $\sqrt{-t} \lesssim 0.6$ GeV/c, indicating that the amplitudes $S_\tau$ and $L_\tau$ are nearly in phase. The phases $\Phi_{US}= \Phi_{U_u}-\Phi_{S_u}$ and $\overline {\Phi}_{US}= \Phi_{U_d}-\Phi_{S_d}$ are also nearly constant and the amplitudes $S_\tau$ and $U_\tau$ are nearly 180$^o$ out of phase. Since the amplitudes $L_\tau$ and $U_\tau$ resonate at $\rho^0(770)$, so must the amplitudes $S_\tau$.\\

The analyses of CERN data on $\pi^+ n \to \pi^+ \pi^- p$ on polarized target at 11.85 and 5.98 GeV/c confirm the expectation of $\rho^0(770)-f_0(980)$ mixing at other energies (reactions 3 and 5 in Table IV.). Figures 12 shows the moduli $|S_u|^2$ and $|S_d|^2$ of the $S$-wave amplitudes at 11.85 at momentum transfers $0.20 \leq -t \leq 0.40$ (GeV/c)$^2$. Despite low statistics, the mass spectra show evidence for $\rho^0(770)$ in the $S$-wave amplitudes also at these energies.\\

There is no $P$-wave and isospin $I=1$ component in the $\pi^0 \pi^0$ states in $\pi^- p \to \pi^0 \pi^0 n$ process. We thus do not expect the $\rho^0(770)$ peak to appear in the $S$-wave intensity in this reaction. The $\rho^0(770)$ interpretation of the observed rho-like resonance in the $S$-wave in $\pi^- p \to \pi^- \pi^+ n$ thus explains the lack of evidence for a narrow rho-like scalar resonance in $\pi^- p \to \pi^0 \pi^0 n$ first observed at CERN in 1972~\cite{apel72} and recently confirmed in high statistics measurements at 18.3 GeV/c at BNL~\cite{gunter01}.

\begin{figure}[ht]
\includegraphics[width=12cm,height=10.5cm]{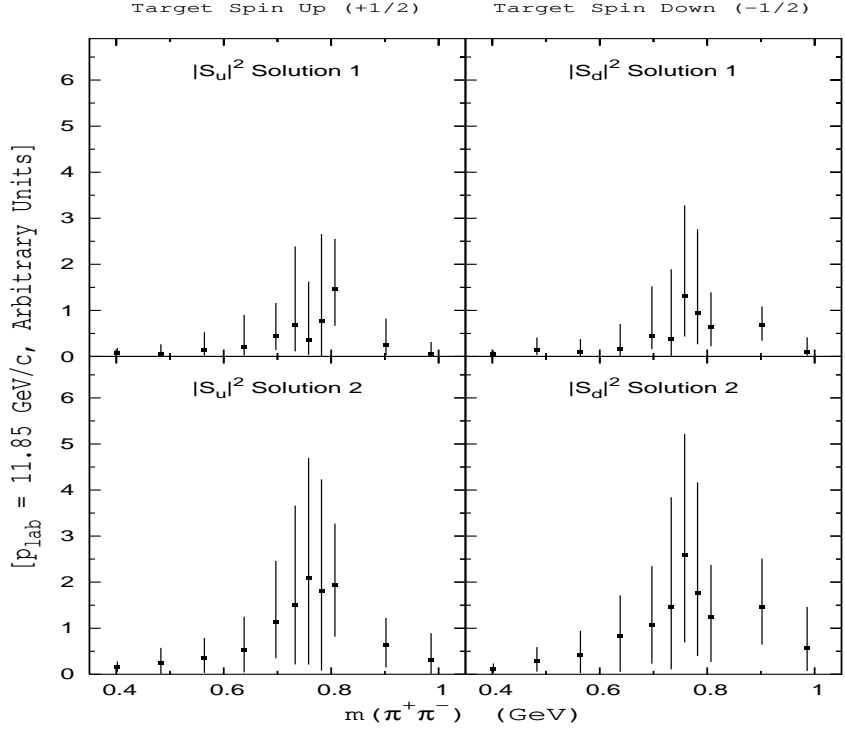}
\caption{Solutions 1 and 2 for the moduli $|S_\tau|^2$, $\tau=u,d$ in $\pi^+ n \to \pi^+ \pi^- p$ at 11.85 GeV/c and $-t$=0.20-0.40 (GeV/c)$^2$.}
\label{Figure 12.}
\end{figure}

\section{Test of rotational and Lorentz symmetry in $\pi^- p \to \pi^- \pi^+ n$.}

It is generally expected that the position and the width of the $\rho^0(770)$ peak as observed in the spin averaged cross-section will be faithfully reproduced on the level of spin amplitudes. The CERN data on polarized target show that this is not the case.\\

From Figures 1,3 and 4 we see that the $\rho^0$ production is suppressed in all target spin "up" amplitudes while the target spin "down" spectra dominate the $\rho^0$ production. The width at half-height of the peaks in the longitudinal spectra  $|L_u|^2$ and $|L_d|^2$ is the expected 155 MeV. However, the $\rho^0$ width shows different values in different transverse spectra. In the spin "down" spectra the $\rho^0$ width is narrower at 120 MeV in $|U_d|^2$ but wider at 180 MeV in $|N_d|^2$. These values are reversed in the spin "up" spectra with 180 MeV and 120 MeV in amplitudes $|U_u|^2$ and $|N_u|^2$, respectively. Such large variations in the $\rho^0$ width are entirely unexpected and appear anomalous.\\

The rotational symmetry of strong interactions requires that the width and mass of a resonance do not depend on its helicity. It also prevents mixing of scalar and vector resonances in the same partial wave with a definite spin $J$. But what if rotational symmetry is broken in pion production? Then the widths of $\rho^0$ in the $P$-wave amplitudes $L_d, U_d$ and $N_d$ could differ and $\rho^0(770)-f_0(980)$ mixing could occur in both $S$- and $P$-wave amplitudes.\\

But how realistic is the hypothesis that rotational invariance is violated in pion production? The breaking of rotational symmetry implies a breaking of Lorentz symmetry. The breaking of Lorentz symmetry has been recently examined theoretically and experimentally in several different contexts. In 2001, Gale and collaborators at McGill examined vector - scalar mixing of $\rho^0(770) - a_0(980)$ resonances in dilepton production in hot and dense matter. The medium-induced breaking of rotational and Lorentz symmetry leads to mixing of different spin states even when the interaction Lagrangian respects all required symmetry properties~\cite{gale01}. While the effects of broken Lorentz symmetry are large at high temperature $T$, they vanish for $T \to 0$. Their work inspired the idea to test rotational and Lorentz symmetry in CERN data on $\pi^- p \to \pi^- \pi^+n$ on polarized targets.\\

The transverse amplitudes $U_\tau$ and $N_\tau$ which show very different $\rho^0$ widths are a mix of dipion +1 and -1 helicities and thus are not suitable to test the rotational symmetry. To test the rotational symmetry we need transversity amplitudes with a definite dipion  helicity . The required amplitudes are $H_u^{+1}, H_u^{-1}, H_d^{+1}, H_d^{-1}$ and they are related to the amplitudes $U_\tau$ and $N_\tau$
\begin{equation}
H^{+1}_u = {1 \over {2}}(U_u + N_u), \quad H^{+1}_d = - {1 \over {2}}(U_d - N_d)
\end{equation}
\[
H^{-1}_u = {1 \over {2}}(U_u - N_u), \quad H^{-1}_d =  + {1 \over {2}}(U_d + N_d)
\]
Their partial wave intensities and polarizations can be calculated from the data on polarized target
\begin{equation}
I(H_u)  = |H_u^{+1}|^2 + |H_u^{-1}|^2 = |U_u|^2 + |N_u|^2
\end{equation}
\[
P(H_u) = -2Re (H_u^{+1}H_u^{-1*} ) = |U_u|^2 - |N_u|^2
\]
The equations for $I(H_d)$ and $P(H_d)$ are similar. For $\rho^0$ with zero helicity we have longitudinal amplitudes $L_u$ and $L_d$. It is convenient to relabel them as
\begin{equation}
H^0_u = L_u, \quad H^0_d = L_d
\end{equation}
Their moduli squared shown in Figure 1 are then the longitudinal intensities 
\begin{equation}
I(H^0_u) = |H^0_u|^2, \quad I(H^0_d) = |H^0_d|^2
\end{equation}

\begin{figure}[h]
\includegraphics[width=12cm,height=10.5cm]{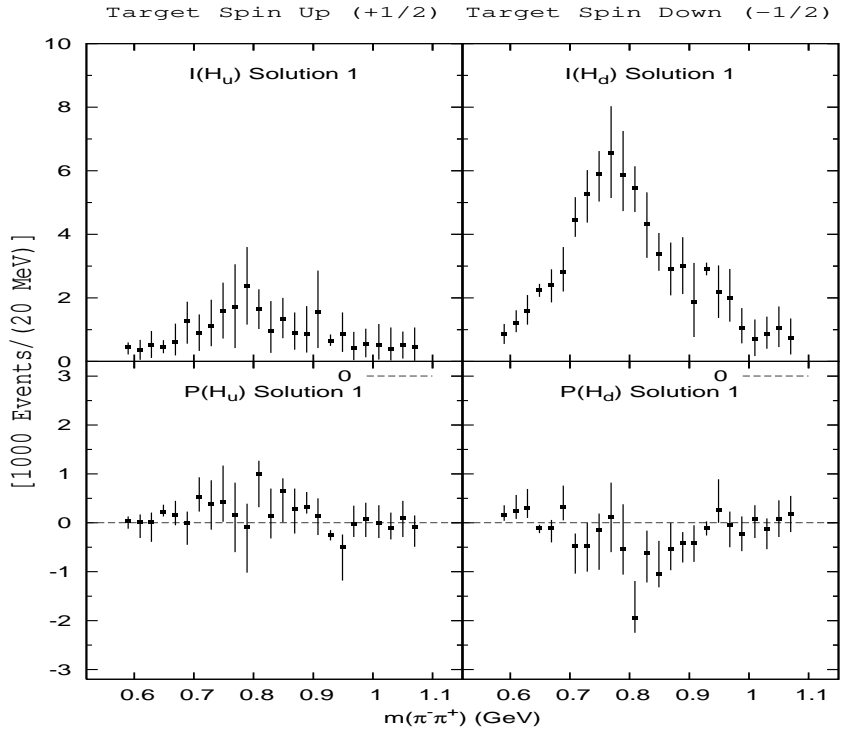}
\caption{Solution 1 for the intensity $I(H_\tau)$ and the interference $P(H_\tau)$, $\tau=u,d$ of the $P$-wave transversity amplitudes $H^{+1}$ and $H^{-1}$ with definite dipion heliciti $\lambda = \pm1$.}
\label{Figure 13.}
\end{figure}

\begin{figure}[h]
\includegraphics[width=12cm,height=10.5cm]{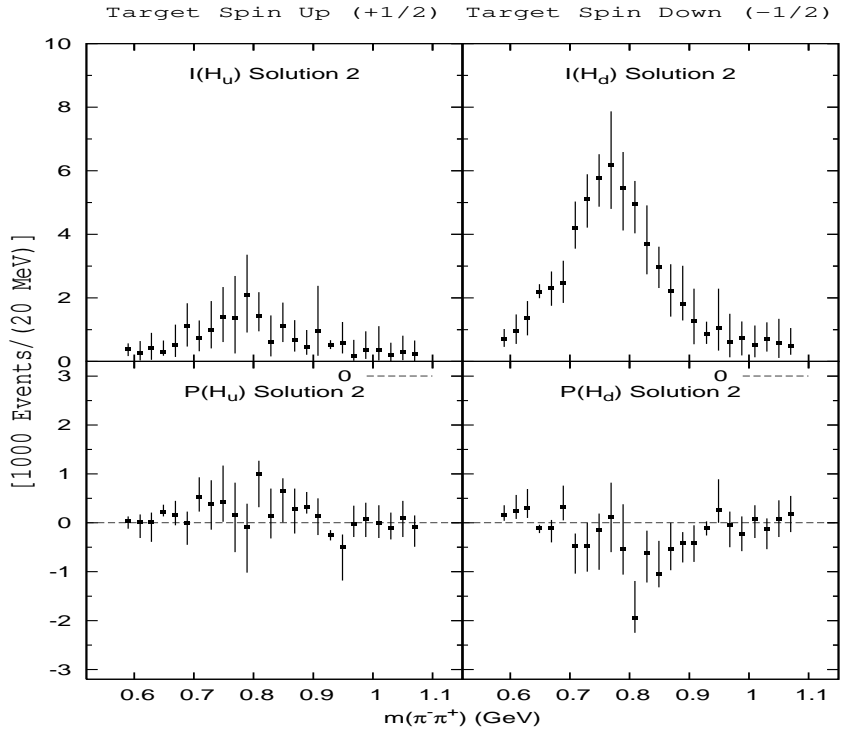}
\caption{Solution 2 for the intensity $I(H_\tau)$ and the interference $P(H_\tau)$, $\tau=u,d$ of the $P$-wave transversity amplitudes $H^{+1}$ and $H^{-1}$ with definite dipion heliciti $\lambda = \pm1$.}
\label{Figure 14.}
\end{figure}

Figures 13 and 14 show the transverse intensities $I(H_u), I(H_d)$ and the polarizations  $P(H_u), P(H_d)$ for the two solutions. The intensities show a clear single peak with the same width at half-height of $\sim$ 155 MeV for both target spins in both solutions. This transverse  $\rho^0$ width is exactly the same as the longitudinal $\rho^0$ width in the intensities $I(H^0_d)$ and $I(H^0_u)$. This indicates that the $\rho^0$ poles in all helicity amplitudes $H^{\lambda}_u, H^{\lambda}_d$,  $\lambda$ = 0, +1, -1 have the same width $\sim$ 155 MeV.\\

This conclusion is supported by the observation that the polarizations $P(H_u)$ and $P(H_d)$ are both small and vanish at the peak mass in $I(H_u)$ and $I(H_d)$, respectively. The pairs of amplitudes $(H^{+1}_u, H^{-1}_u)$ and $(H^{+1}_d, H^{-1}_d)$ with opposite transverse helicities are $90^{\circ}$ out of phase at the peak mass and move around $90^{\circ}$ at other masses. Such behaviour is consistent with the two amplitudes in each pair being dominated by the same $\rho^0$ Breit-Wigner pole interfering with a weak non-resonating background.\\

On this basis we can conclude that there is no evidence for any dependence of $\rho^0$ width on its helicity from the CERN data on pion production on polarized target. To explain the large differences in the $\rho^0$ witdth observed in the spectra $|U_d|^2$ and $|N_d|^2$ we note from (7.2)
\begin{equation}
|U_d|^2 = {1 \over {2}}( I(H_d) + P(H_d) )
\end{equation}
\[
|N_d|^2 = {1 \over {2}}( I(H_d) - P(H_d) )
\]
Figures 13 and 14 show that the polarization $P(H_d)$ is negative and broad in the $\rho^0$ mass region. As the result of the opposite signs of $P(H_d)$ in (7.5), the $\rho^0$ width is narrower in $|U_d|^2$ and wider in $|N_d|^2$. This apparent difference is entirely due to the interference of the amplitudes $H^{+1}_d$ and $H^{-1}_d$. This interference arises from the presence of a non-resonating background (continuum) in these amplitudes. The relations for the target spin "up" spectra $|U_u|^2$ and $|N_u|^2$ have the same form as Eqs. (7.5). However, now the polarization $P(H_u)$ is positive and broad around the $\rho^0$ mass and it has therefore an opposite effect on their widths, as observed.\\

Rotational invariance of strong interactions and the particle interpretation of resonances also require that the $\rho^0$ mass $m_\rho$ does not depend on its helicity $\lambda$. The test of this aspect of rotational symmetry in pion production is somewhat less direct since the positions of the observed peaks in the $P$-wave spectra need not correspond exactly to the $\rho^0$ mass as the result of interference effects with non-resonating background.\\

Figures 13 and 14 show that the transverse intensities $I(H_d)$ and $I(H_u)$ for target spins "down" and "up" peak at 770 and 790 MeV, respectively. In Figures 1 we see that the longitudinal intensities $I(H^0_d)$ and $I(H^0_u)$ for target spins "down" and "up" peak at 790 and 810 MeV, respectively. The apparent dependence of the peak positions on the target spin can be understood as the intereference of the $\rho^0$ pole with the target spin dependent background. We may conclude that there is no evidence for the dependence of the $\rho^0$ mass on its helicity from the CERN data on pion production on polarized target.\\

The amplitude analysis of $\pi^- p \to \pi^- \pi^+ n$ on polarized target presented in Section III. provides the first direct experimental evidence for the independence of the mass and width of a resonance on its helicity expected theoretically from the rotational symmetry of strong interactions. If strong interactions do not violate Lorentz symmetry in pion production then we have to look somewhere else to find an explanation for the observed $\rho^0(770) - f_0(980)$ mixing.

\section{Non-existence of a unique analytical solution and\\ the Central Hypothesis.}

Apart from a few Monte Carlo samplings in one $m$-bin in reaction 1 and in 4 $t$-bins in reaction 2, all Monte Carlo samplings of the data error volume that yield physical solutions for the moduli and cosines require that $R>0$. This numerical result means that the two solutions for the moduli $|L_\tau|^2$ have analytical form
\begin{eqnarray}
|L_\tau(1)|^2 & = & -y_0+R \cos ({\phi \over {3}}) + \sqrt{3} R \sin({\phi \over {3}})\\
\nonumber
|L_\tau(2)|^2 & = & -y_0+R \cos ({\phi \over {3}}) - \sqrt{3} R \sin({\phi \over {3}})
\nonumber
\end{eqnarray}
where we suppressed the subscript $\tau$ on r.h.s. of (8.1). It follows from (2.17) and (6.1) that the two solutions for the moduli $|S_\tau|^2$, $|U_\tau|^2$ and $|N_\tau|^2$ have a similar analytical form.\\

From the Table III. we see that the condition $Q^2+P^3=0$ defines a surface $\Sigma_{0,\tau}$ which divides the error volumes of observables $a_k$, and $\overline {a}_k$, $k=1,6$ into two separate regions $\Sigma_{+,\tau}$ and $\Sigma_{-,\tau}$ with $Q^2+P^3>0$ and $Q^2+P^3<0$, respectively. In $\Sigma_{+,\tau}$ there are two complex conjugate solutions and a negative real solution for $|L_\tau|^2$. In $\Sigma_{-,\tau}$ there are two distinct positive solutions and one negative solution for $|L_\tau|^2$. On the boundary $\Sigma_{0,\tau}$ there are two degenerate solutions corresponding to $\sin({\phi \over {3}})=0$ and comprising one positive solution, and one negative solution. The two conditions $Q^2+P^3=0$ for the observables $a_k$, and $\overline {a}_k$, $k=1,6$ divide the data error volume of the measured density matrix elements similarly into two regions $\Sigma_+$ and $\Sigma_-$ characterized by complex conjugate and positive solutions for the moduli, respectively.\\

From the standard Quantum Field Theory we expect a unique physical solution. This expectation raises a question whether such unique physical solution could be identified with the single  positive solution on the boundary $\Sigma _0$. This solution would describe a unitary evolution from pure initial states to pure final states, would be time-reversible and would represent the pion creation process as an isolated event in the Universe. \\

We investigated this possibility numerically as follows. The difference between the solutions, which we call level splitting,  
\begin{equation}
\delta_\tau=|L_\tau(1)|^2-|L_\tau(2)|^2=2 \sqrt{3} R_\tau \sin ( {\phi_\tau \over {3}})
\end{equation}
measures the distance of the solutions from the boundary $\Sigma_{0,\tau}$. We looked at a number of physical solutions constrained progressively closer to the boundary by imposing a condition on the solutions that $\delta_\tau < \delta$ where the upper bound $\delta$ was progressively reduced. The amplitude analysis was done with normalized amplitudes such that
\begin{equation}
{d^2 \sigma \over {dm dt}}=\sum \limits_{A=S,L,U,N} |A_u|^2+|A_d|^2 =1
\end{equation}
For each limit $\delta$ the analysis was done with 100 million Monte Carlo samplings of the data error volume. The results are shown in Table V. \\

\begin{table}
\caption{Total number of physical solutions in progressively closer vicinities of the boundary $\Sigma_0$. The parameter $\delta$ is upper limit on level splitting for normalized amplitudes $|L_\tau|^2, \tau=u,d$. For $\delta=free$ results were scaled from sample size $5.10^6$ to $100.10^6$. }
\begin{tabular}{cccccc}
\toprule
$\delta$ &  $free$  &  0.0010  &  0.0005  &  0.00025  &  0.0001  \\
\colrule
Sample size  &  $20(5.10^6)$  &  $100.10^6$  &  $100.10^6$  &  $100.10^6$  &  $100.10^6$  \\
Physical solutions  &  66008240  &  12499  &  1075  &  82  &  2 \\
\botrule
\end{tabular}
\label{Table V.}
\end{table}

The results clearly show that the number of physical solutions dramatically decreases as the boundary is approached. The real solutions on the boubary are thus not physical. Physical solutions are confined to a region away from the boundary leading to two distinct solutions for the moduli $|A(i)|^2 \equiv |A_u(i)|^2$ and $|\overline {A}(j)|^2 \equiv |A_d(j)|^2$, $i,j=1,2$. This multiplicity of solutins is inherent in the data in all 8 analyzed reactions  listed in the Table IV.\\

We have shown in Section V. that the complete measurements on polarized target uniquely determine the phases of reduced $S$- and $P$-wave transversity amplitudes $A$ and $\overline {A}$, $A=S,L,U,N$. In a sequel paper~\cite{svec07c} we shall show that the relative phase $\omega$ in the $S$- and $P$-wave transversity amplitudes $A_\tau$ is uniquely determined by a process of conversion of transversity amplitudes into the helicity amplitudes. Up to an inessential overall phase, the $S$- and $P$-wave transversity amplitudes $A_u(i)$ and $A_d(j)$ are thus uniquely determined for each of the four solutions $i,j=1,2$.\\

The four solutions cannot be distinguished in measuremnts of reduced density matrix $\rho^0_f$ on polarized targets since for all measured density matrix elements
$\rho^0_f = \rho^0_f(ij)$ for all $i,j$. However the predictions for full density matrix, and thus for the measurable recoil nucleon polarizations, are all different $\rho_f(ij)$. We could hope that the measurements of recoil nucleon polarization would select one valid physical solution. But we are also free to think that all four solutions are valid physical solutions and that their multitude reflects new and important aspects of reality.\\
 
This is our central hypothesis. It implies that the measured final state density matrix is a mixed state of the four physical solutions 
\begin{equation}
\rho_f(\theta \phi,\vec {P})=p_{11} \rho_f(11,\theta \phi, \vec {P})+p_{12} \rho_f(12, \theta \phi,\vec {P})+p_{21} \rho_f(21, \theta \phi,\vec {P}) +p_{22} \rho_f(22, \theta \phi,\vec {P})
\end{equation}
where the probabilities $p_{ij}$ satisfy condition $\sum \limits_{i,j=1,2} p_{ij} =1$
and $\vec {P}$ is the target polarization. In a sequel paper~\cite{svec07c} we show that each set $A_u(i),A_d(j),i,j=1,2$ of transversity amplitudes uniquely determines the $S$- and $P$-wave subsystem of the density matrix $\rho_f(ij,\theta \phi, \vec{P})$. The transversity amplitudes for higher spins $J$ also come in four sets $A^J_{\lambda,u}(i),A^J_{\lambda,d}(j),i,j=1,2$ so that the corresponding density matrix $\rho_f(ij,\theta \phi, \vec{P})$ is fully determined. The probabilities $p_{ij}$ can be experimentally determined in measurements of recoil nucleon or hyperon polarization on polarized target~\cite{svec07c}.\\

The hypothesis (8.4) implies that pure initial states evolve into mixed final states in the pion creation process $\pi^- p \to \pi^- \pi^+n$. In our previous paper ~\cite{svec07a} we have shown how the conditions for unitary evolution of pure initial states into pure final states are violated by data in $\pi^- p \to \pi^- \pi^+ n$, implying a non-unitary evolution in pion creation process. In quantum theory such non-unitary evolution arises in interactions of open quantum systems with an environment~\cite{kraus83,nielsen00,breuer02}. Physically the hypothesis (8.4) thus means that the pion creation process can be thought of as open quantum system interacting with a quantum environment. Such interaction is described by a unitary co-evolution operator $U$ replacing the standard $S$-matrix.\\

In the next Section IX. we show how the analytical solutions for the transversity amplitudes reveal the existence of a new quantum number $g$ characterizing the quantum states of the environment. In Section X. we show that these analytical solutions can be identified with co-evolution amplitudes which are matrix elements of the co-evolution operator $U$ corresponding to the quantum states of the environment. 

\section{Level splitting and the quantum number $g$.} 

When we look at the mass spectra of the amplitudes $|S|^2, |L|^2, |U|^2$ and $|N|^2$ in Figures 1-4 we notice that the two solutions are very close. The spectra carry definite quantum numbers of the target nucleon transversity corresponding to a two-level quantum system. They also carry the angular quantum numbers $J, \lambda$ of the two-pion state corresponding to one-level and three-level quantum systems for the $S$- and $P$-wave spectra, respectively. If we think of the mass spectra as variable spectral lines in dipion mass, the two solutions suggest a level splitting due to some unknown interaction associated with a new two-valued quantum number $g= \pm 1$ corresponding to the signs $\pm 1$ in the analytical solutions (8.1). From the hypothesis (8.4) it follows that the level splitting arises from the interaction of the pion creation process with a quantum environment.\\

The difference between the two solutions for the $P$-wave amplitudes has the same value
\begin{equation}
\Delta_\tau = |L_\tau(1)|^2 - |L_\tau(2)|^2
\end{equation}
indicating that the level splitting does not depend on the dipion helicity $\lambda$. It depends on the spin J and for the $S$-wave 
\begin{equation}
|S_\tau(1)|^2 - |S_\tau(2)|^2 = - 3 \Delta_\tau
\end{equation}
The results for $\Delta_\tau$ are shown in Figure 15. The largest level splitting occurs in $\Delta_d$ for masses in the vicinity of the $f_0(980)$ resonance.\\

\begin{figure}[ht]
\includegraphics[width=12cm,height=10.5cm]{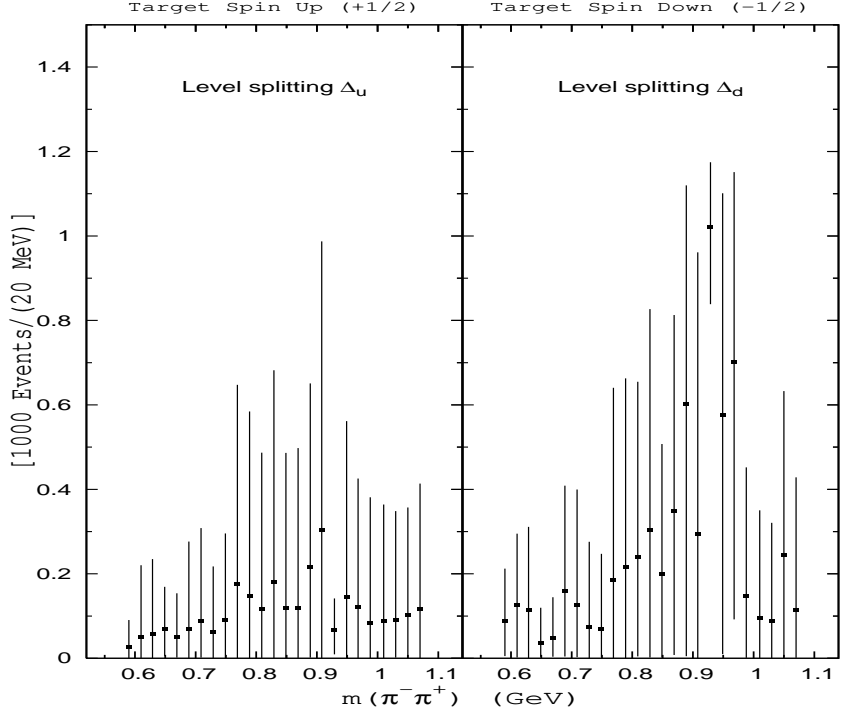}
\caption{Level splitting $\Delta_\tau$, i.e. the difference between the Solution 1 and Solution 2 for the unnormalized moduli squared $|L_\tau|^2, \tau=u,d$.}
\label{Figure 15.}
\end{figure}

We see from (8.1) that the $P$-wave moduli have a general form
\begin{eqnarray}
|A_\tau(1)|^2 & = & (X_{A \tau})^2+(Y_\tau)^2\\
\nonumber
|A_\tau(2)|^2 & = & (X_{A \tau})^2-(Y_\tau)^2
\nonumber
\end{eqnarray}
where $(Y_\tau)^2={1 \over {2}} \Delta_\tau$. Next we define amplitudes associated with the quantum number $g_\tau$
\begin{equation}
Z_{A \tau} (g_\tau) = \exp(ig_\tau \lambda_{A \tau})|A_\tau(1)|
\end{equation}
where the phase $\lambda_{A \tau}$ is given by the level splitting 
\begin{equation}
\tan \lambda_{A \tau} = {Y_\tau \over {X_{A \tau}}} 
\end{equation}
The intensities (9.3) then read as interferences
\begin{eqnarray}
|A_\tau(1)|^2 & = & Re \bigl ( Z_{A \tau}(+1)Z_{A \tau}^*(+1) \bigr )\\ 
\nonumber
|A_\tau(2)|^2 & = & Re \bigl ( Z_{A \tau}(+1)Z_{A \tau}^*(-1) \bigr )
\nonumber
\end{eqnarray}
In the first interference term the phase difference $(+1) \lambda_{A \tau} - (+1) \lambda_{A \tau} =0$. The phase difference in the second interference term is non-zero $(+1) \lambda_{A \tau} - (-1) \lambda_{A \tau} = 2(+1) \lambda_{A \tau}$ so that
\begin{equation}
|A_\tau(2)|^2 =\cos (2 \lambda_{A \tau}) |A_\tau(1)|^2
\end{equation}
Next we associate the solutions for the $P$-wave amplitudes with the quantum number $g_\tau$
\begin{equation}
A_\tau(1) \equiv A_\tau (+1)= |A_\tau(1)| \exp i\Phi_{A \tau}(1) = \exp(+i \lambda_{A \tau}) |A_\tau (1)| \exp( i\Psi_{A \tau} (1) 
\end{equation}
\[
A_\tau(2) \equiv A_\tau (-1)= |A_\tau(2)| \exp i\Phi_{A \tau}(2) = \exp(-i \lambda_{A \tau}) |A_\tau (2)| \exp( i\Psi_{A \tau}(2))
\] 
With $\Psi_{A \tau}(+1) \equiv \Psi_{A \tau}(1)$ and $\Psi_{A \tau}(-1) \equiv \Psi_{A \tau}(2)$ we can write the amplitudes $A_\tau(g)$ in the form
\begin{equation}
A_\tau(+1)=Z_{A \tau} (+1) \exp (i \Psi_{A \tau}(+1))
\end{equation}
\[
A_\tau(-1)=\sqrt{\cos (2 \lambda_{A \tau})} Z_{A \tau} (-1) \exp (i \Psi_{A \tau}(-1))
\]
\[
=\sqrt{\cos (2 \lambda_{A \tau})} \exp(-2i \lambda_{A \tau})Z_{A \tau} (+1) \exp (i \Psi_{A \tau}(-1))
\]
With these expressions we find
\begin{equation}
|A_\tau(+1)|^2  =  Re (A_\tau(+1) A_\tau^*(+1))=Re (Z_{A \tau} (+1)Z_{A \tau}^* (+1))
\end{equation}
\[
|A_\tau(-1)|^2  =  Re (A_\tau(-1) A_\tau^*(-1))=Re (Z_{A \tau} (+1)Z_{A \tau}^* (-1))
\]
For the $S$-wave we have $|S_\tau(1)|^2 \leq |S_\tau (2)|^2$ so we have to define
\begin{equation}
Z_{S \tau}(g_\tau)=\exp(ig_\tau \lambda_{S \tau}) |S_\tau(2)|
\end{equation}
Then the moduli are interferences
\begin{eqnarray}
|S_\tau(1)|^2 & = & Re ( Z_{S \tau}(-1)Z_{S \tau}^*(+1) )\\
\nonumber
|S_\tau(2)|^2 & = & Re ( Z_{S \tau}(-1)Z_{S \tau}^*(-1) )
\nonumber
\end{eqnarray}
and the amplitudes associated with quantum number $g_\tau$ read
\begin{equation}
S_\tau(1) \equiv S_\tau (+1)= \sqrt{\cos (2 \lambda_{S \tau})}Z_{S \tau}(+1) \exp(i\Psi_{S \tau}(+1))
\end{equation}
\[ 
S_\tau(2) \equiv S_\tau (-1)= Z_{S \tau}(-1) \exp(i\Psi_{S \tau}(-1))
\]
where $\Psi_{S \tau}(+1) \equiv \Psi_{S \tau}(1)$ and $\Psi_{S \tau}(-1) \equiv \Psi_{S \tau}(2)$.\\

\begin{figure}[ht]
\includegraphics[width=12cm,height=10.5cm]{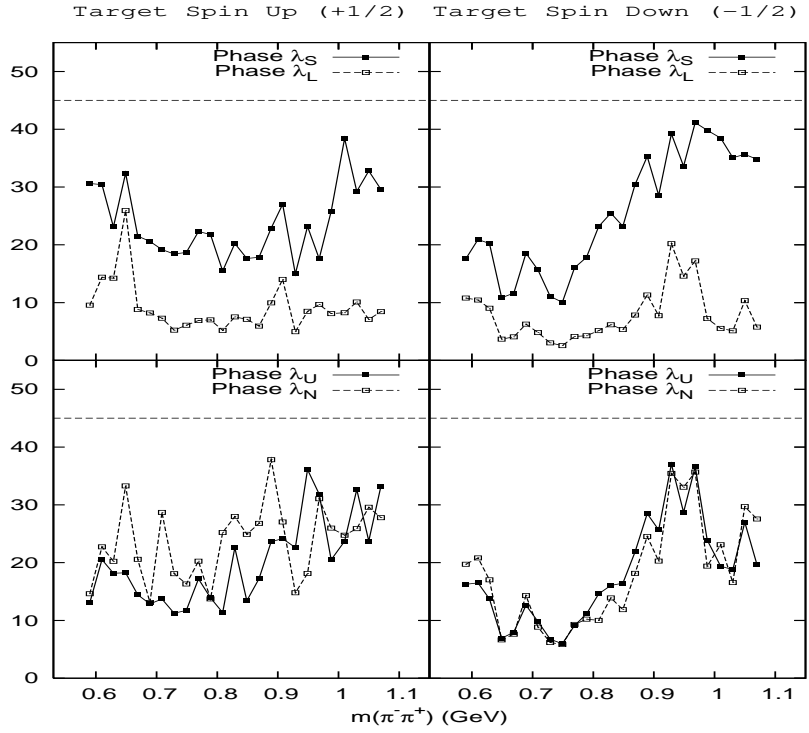}
\caption{Level splitting phases $\lambda_{A \tau}$ for amplitudes $A=S,L,U,N$. Maximum value of the phases is 45$^o$.}
\label{Figure 16.}
\end{figure}
 
The level splitting phases $\lambda_{A \tau}$ are shown in Figure 16. From the definituion  (9.5) we see that the phase $\lambda_{A \tau}$ describes the relative strength of the interaction with the environment in the amplitude $A_\tau$.  For small $\lambda_{A \tau}$ the interaction has a small relative effect. The effect is maximum at 45$^o$. when $|A_\tau(2)|^2=0$. The interaction has a larger relative effect in the $S$-wave amplitudes $S_\tau$ than in the $P$-wave amplitudes $L_\tau$ with zero dipion helicity. The effect is about the same in the transverse amplitudes $U_\tau$ and $N_\tau$. The largest overall effect is in the $A_d$ amplitudes near the $f_0(980)$ resonance mass. A characteristic feature of all level splitting phases are phase fluctuations.

\section{Quantum states of the environment.}

In order to introduce the central concept of co-evolution amplitudes we first briefly review the Kraus representation for for reduced density matrices of open quantum systems interacting with an environment. It is the co-evolution amplitudes which involve the interacting degrees of freedom of the environment and with which we shall identify the solutions for the transversity amplitudes.\\

The co-evolution of an open quantum system $S$ with a quantum environment $E$ is a unitary evolution~\cite{nielsen00}
\begin{equation}
\rho_f(S,E)=U \rho_i(S,E)U^+ =U \rho_i(S) \otimes \rho_i(E) U^+
\end{equation}
The initial state of the environment is in general a mixed state
\begin{equation}
\rho_i(E)= \sum \limits_{\ell} p_{\ell \ell'} |e_\ell><e_{\ell'}|
\end{equation}
where $|e_\ell>$ are quantum states of interacting degrees of the environment and $\sum \limits_\ell p_{\ell \ell}=1$. The Hilbert space of the environment has a finite dimension. It is given by a condition $\dim {H(E)} \leq \dim{H_i(S)} \dim{H_f(S)}$~\cite{nielsen00}. After the interaction the system $S$ is fully described by reduced density matrix given by Kraus representation
\begin{equation}
\rho_f(S)=Tr_E(\rho_f(S,E))= \sum \limits_\ell \sum \limits_{m,n} 
p_{mn} S_{\ell m} \rho_i(S) S_{n \ell}^+
\end{equation}
where the operators $S_{\ell m}=<e_\ell|U|e_m>$ satisfy a completness relation 
$\sum \limits_\ell \sum \limits_{m,n} S_{n \ell}^+ S_{\ell m}=I$.\\

In our next step we associate the two solutions for transversity amplitudes $A_u(i)$ and $A_d(j)$, $i,j=1,2$ with two single qubit states $|i>$ and $|j>$, respectively.
Then the hypohesis (8.4) allows us to identify the four degrees of freedom of the environment $|e_\ell>$ allowed by the condition  
$\dim {H(E)} \leq \dim{H_i(S)} \dim{H_f(S)}=(2s_p+1)(2s_n+1)=4$ with the four two-qubit states  $|e_\ell>=|i>|j>$. Since the transversity amplitudes can possess only one solution at a time, the co-evolution amplitudes $A^{J \eta}_{\lambda, \tau}(\ell m)$ 
\begin{equation}
A^{J \eta}_{\lambda, \tau}(\ell m)=<J \lambda \eta,\tau_n|<e_\ell|U|e_m>|0 \tau_p>
\end{equation}
must be diagonal for any dipion spin $J$ and naturality $\eta$
\begin{equation}
A^{J\eta}_{\lambda, \tau}(\ell m)=A^{J\eta}_{\lambda, \tau}(\ell \ell)\delta_{\ell m}=
A^{J\eta}_{\lambda, \tau}(ij,ij) \delta_{ij,i'j'} \equiv
A^{J\eta}_{\lambda, \tau}(ij) \delta_{ij,i'j'} 
\end{equation}
where
\begin{equation}
A^{J\eta}_{\lambda, u}(ij)=
<J\lambda \eta,\tau_n|<ij|U|ij>|0 u>=A^{J \eta}_{\lambda, u}(i)
\end{equation}
\[
A^{J\eta}_{\lambda, d}(ij)=
<J\lambda \eta,\tau_n|<ij|U|ij>|0 d>=A^{J \eta}_{\lambda, d}(j)
\]
In (10.4)-(10.6) the state $|0 \tau_p>$ is the $\pi^- p$ initial state with pion spin 0 and proton transversity $\tau_p=u,d$. In (10.6) $A^{J \eta}_{\lambda, u}(i),
A^{J \eta}_{\lambda, d}(j),i,j=1,2$ are solutions for transversity amplitudes with dipion spin $J$ and helicity $\lambda$ generalizing the $S$- and $P$-wave amplitudes $A_u(i),A_d(j)$.\\

Instead of using the solution qubits $|i>$ and $|j>$ to define the quantum states $|i>|j>$ of the environment and the co-evolution amplitudes $A^{J \eta}_{\lambda, \tau}(ij)$, we could have used the qubits $|g_u>$ and $|g_d>$ to define equivalent states $|g_u>|g_d>$ of the environment and the equivalent co-evolution amplitudes $A^{J \eta}_{\lambda, \tau} (g_u g_d) \equiv A^{J \eta}_{\lambda, \tau}(g_\tau)$. These states more closely reflect the qubit nature of the interacting degrees of the environment and may possess a deeper physical meaning. Recall that the solutions for the moduli are distinct and well away from the boundary  $\Sigma_0$. This suggests that the quantum numbers $g_\tau$ are good quantum numbers suitable to describe the interacting degrees of freedom of the environment.\\

\section{Non-unitary dynamics of $\rho^0(770) - f_0(980)$ mixing.}

Non-unitary evolution to mixed final states (8.4) in $\pi^- p \to \pi^- \pi^+ n$ arises from a $CPT$ violating interaction of the pion creation process with a quantum environment. The hypothesis of the existence of such an environment explains naturally the observed $\rho^0(770)-f_0(980)$ mixing. We can see how this may work using the following mechanical analogy. Consider a pendulum of mass $m$ and length $L$ oscillating with a natural frequency $\omega$, and an ensemble of several other penduli with various masses $m_i$ and lengths $L_i$. When the other penduli are isolated from the oscillating pendulum they do not oscillate. However, when all the penduli are attached to a common rod all penduli begin to oscillate with the same resonant frequency although with different amplitudes. The interaction of the penduli with a common environment - the rod - allows a resonance from one pendulum to "leak" into the other penduli which have different natural frequencies. Similarly we can imagine that the resonant $q \overline{q}$ modes produced in $\pi^- p \to \pi^- \pi^+ n$ process interact with an environment which allows resonances to "leak" into different two-pion partial wave amplitudes.\\

The requirement that the Kraus representation leaves invariant the spin formalism used in the data analysis necessitates~\cite{svec07a} that the co-evolution amplitudes
\begin{equation}
A^{J \eta}_{\lambda, \tau}(g_\tau)=<J \lambda \eta,\tau_n|<g_ug_d|U|g_ug_d>|0 \tau>
\end{equation}
transform under $P$-parity as a two-body $P$-parity conserving process $\pi^- + p \to "J(\pi^- \pi^+)" + n$ with parity $P=(-1)^J$ for the dipion states $"J(\pi^- \pi^+)"$. This means that there is no vector associated with the quantum states $|g_ug_d>$ of the environment. In particular, there is no energy-momentum associated with these quantum states. The interacting hadrons conserve their energy-momentum and there is no exchange of energy-momentum with the environment, in agreement with the original proposal by Hawking for particle processes interacting with quantum fluctuations of the space-time
 metric~\cite{hawking82}. Instead, the interaction with the environment is a 
non-dissipative dephasing process. The co-evolution amplitudes $U_{fi}$ can then be written in a form
\begin{equation}
U_{fi}=I_{fi}+i(2 \pi)^4 \delta^4(P_f-P_i)T_{fi}
\end{equation}
where $P_i$ and $P_f$ are total four-momenta of the initial and final hadron states and $T_{fi}$ is the transition matrix for the process $|\pi^-p>+|g_ug_d> \to |\pi^- \pi^+ n>+|g_ug_d>$.\\ 

To construct a model of interaction of the pion creation process $S$ with the environment $E$ we can think of the transition operator $T$ as a product of three evolution operators 
\begin{equation}
T=T_{fE}T_PT_i
\end{equation}
corresponding to three different stages of the pion creation process. In the first stage the operator $T_i$ maps the initial hadron state $|i>=|0 \tau>$ from Hilbert space $H_i(\pi^-p)$ to a vector $|i_{int}>$ in an intermediate (hidden) Hilbert space $H_{int}(q \overline{q}n)$ of resonant and non-resonant $(q \overline{q})$ modes and neutron states which form an orthonormal basis in $H_{int}(q \overline{q}n)$. In general, these states could include formations also of four-quark states such as diquark-antidiquaurk states. For simplicity we assume that the resonant modes correspond to quark-antiquark states in $LS$-coupling. Focusing on the resonant modes component of $|i_{int}>$ we have
\begin{equation}
T_i|i>|g_ug_d>=|i_{int}>|g_ug_d>= 
\end{equation}
\[
\Bigl (\sum \limits_{K \mu} <K \mu(q \overline{q}) \eta, I_K, \tau_{K \mu}|T_i|0 \tau>
|K \mu(q \overline{q}), I_K, \tau_{K \mu}> \Bigr )|g_ug_d>
\]
where $K$ and $\mu$ are the spin and helicity of the $q \overline{q}$ state, $I_K$ is its isospin, $\tau_{K \mu}$ is the transversity of the recoil neutron in the intermediate process $\pi^- p \to q \overline{q} n$ and $\eta$ is its $t$-channel naturality. The resonant states propagate with Breit-Wigner amplitudes $a_K(m^2)$ and the coherent state $|i_{int}>$ is modified in $H_{int}(q \overline{q}n)$
by the action of the propagation operator $T_P$
\begin{equation}
|i_{int,P}>|g_ug_d>=\bigl (T_P|i_{int}> \bigr ) |g_ug_d>=
\end{equation}
\[ 
\Bigl (\sum \limits_{K \mu} <K \mu(q \overline{q}) \eta, I_K, \tau_{K \mu}|T_i|0 \tau>
T_P|K \mu(q \overline{q}), I_K, \tau_{K \mu}> \Bigr )|g_ug_d>=
\]
\[
\Bigl (\sum \limits_{K \mu} <K \mu(q \overline{q}) \eta, I_K, \tau_{K \mu}|T_i|0 \tau>
a_K(m^2)|K \mu(q \overline{q}), I_K, \tau_{K \mu}> \Bigr )|g_ug_d>
\]

As the resonant modes propagate they interact with the environment. It is crucial to realize that the interaction involves simultaneosly the whole coherent state $|i_{int,P}>$. In this final stage the the states $|K \mu(q \overline{q}), I_K, \tau_{K \mu}>|g_ug_d>$ are mapped into the Hilbert space $H_f(\pi^- \pi^+ n) \otimes H(E)$
\begin{equation}
T_{fE}|K \mu(q \overline{q}), I_K,\tau_{K \mu}>|g_ug_d>=
\end{equation}
\[
\sum \limits_{J \lambda} <g_ug_d|<J \lambda(\pi^- \pi^+),I_K, \tau_n|T_{fE}
|K \mu(q \overline{q}),I_K,\tau_{K \mu}>|g_ug_d>|J \lambda(\pi^- \pi^+),I_K,\tau_n>|g_ug_d>
\]
where we have assumed isospin conservation in transitions from $|K\mu(q\overline{q}),I_K,\tau_{K\mu}>$ to $|J\lambda(\pi^-\pi^+),I_K,\tau_n>$. The total co-evolution amplitude will involve isospin amplitudes
\begin{equation}
<J\lambda(\pi^- \pi^+) \eta, I_K, \tau_n|<g_ug_d|T|g_ug_d>|0 \tau>=
\end{equation}
\[
\sum \limits_{K \mu}<J\lambda(\pi^- \pi^+) \eta, I_K, \tau_n|<g_ug_d|T_{fE}|g_ug_d>|K \mu(q \overline{q}),I_K, \tau_{K\mu}>
\]
\[
 a_K(m^2)<K \mu(q \overline{q}),I_K, \tau_{K\mu}|T_i|0 \tau>
\]
where $I_K=I(\pi^-\pi^+)=0,1$ is the isospin of the $\pi^-\pi^+$ state. The non-resonant modes with $I_K=2$ give rise to a non-resonating amplitude 
$<J \lambda(\pi^-\pi^+),I(\pi^-\pi^+)=2,\tau_n|<g_ug_d|T|g_ug_d>|0\tau>$ given by an expression similar to (11.7) with a replacements $a_K(m^2) \to 1$ and $q\overline{q} \to q\overline{q}q\overline{q}$.\\

In standard spectroscopy there is no environment interaction and the matrix elements of $T_{fE}$ are diagonal with $J(\pi^- \pi^+)=K(q \overline{q})$ and 
$\lambda(\pi^- \pi^+)=\mu(q \overline{q})$. As a result, there is no mixing of resonances in two-pion partial wave amplitudes as the basis vectors in $H_{int}(q \overline {q}n)$ are mapped into basis vectors in $H_f(\pi^- \pi^+n)$. The effect of the interaction of the resonant modes with the environment are transitions from the resonant modes 
$|K \mu(q \overline{q},I_K,\tau_{K \mu}>$ to two-pion states 
$|J \lambda(\pi^- \pi^+),I_K,\tau_n>$ as the basis states in $H_{int}(q \overline {q}n)$ are mapped into superpositions of basis vectors in $H_f(\pi^- \pi^+n)$. The non-diagonal matrix   matrix elements of $T_{fE}$ allow for the resonances to "leak" into different two-pion partial wave amplitudes and account for the observed $\rho^0(770)-f_0(980)$ mixing. The interaction with the environment is non-local and $CPT$ violating which accounts for the violation of the conservation of angular momentum and parity in the non-diagonal transitions between these two different quantum systems~\cite{greenberg02}.

\section{Entanglement of $\pi^- \pi^+$ isospin states and $CPT$ violation.}

Before we can write down the final form of the co-evolution amplitudes we need to clarify the isospin structure of the produced two-pion states. The Generalized Bose-Einstein symmetry 
assumes that all particles in an isospin multiplet are identical particles. This symmetry implies that spin and isospin of two-pion non-interacting states must satisfy condition  $J+I$ = even~\cite{martin70}. As the result, the symmetrized $\pi^- \pi^+$ isospin state is symmetric for $J=even$ and antisymmetric for $J=odd$
\begin{eqnarray}
\text{J=even} \quad |S> & = & {1 \over{\sqrt{2}}}(|\pi^->| \pi^+>+|\pi^+>|\pi^->)=
-{1 \over {\sqrt{3}}}(\sqrt{2}|0,0>+|2,0>)\\
\nonumber
\text{J=odd} \quad |A> & = & {1 \over{\sqrt{2}}}(|\pi^->| \pi^+>-|\pi^+>|\pi^->) = |1,0>
\nonumber
\end{eqnarray}
where we used the convention $|\pi^+>=-|1,+1>$~\cite{gibson76} and replaced the symmetrization normalization factor ${1 \over{2}}$ by ${1 \over {\sqrt{2}}}$. The states $|S>$ and $|A>$ then acquire a meaning of maximally entangled Bell states of two-pion charge states with pion charge states $|\pi^->$ and $\pi^+>$ representing the two qubit states.\\

The interaction with the environment is assumed to conserve the isospin $I_K$ and $G$-parity. The $G$-parity of $n$ pion states is $G=(-1)^n$ while isospin multiplet of particle-antiparticle pairs in $LS$ states has $G=(-1)^{L+S+I}$~\cite{martin70}. In $\pi^- p \to \pi^- \pi^+ n$ the $q \overline {q}$ and $q^2 \overline{q}^2$ resonant modes must be in triplet states of the quark and diquark spins~\cite{flamm82}. The conservation of $G$-parity then implies that for resonant modes $K+I_K$ = even. For states with $J$=even (odd) the transitions from $K$=even (odd) will conserve Generalized Bose-Einstein symmetry while the transitions from $K$=odd (even) will violate it. We thus can introduce Bose-Einstein symmetry conserving and violating two-pion isospin states
\begin{equation}
|I_C(\pi^- \pi^+)>={1 \over{\sqrt{2}}}(|\pi^->| \pi^+>+(-1)^J|\pi^+>|\pi^->)
\end{equation}
\[
|I_V(\pi^- \pi^+)>={1 \over{\sqrt{2}}}(|\pi^->| \pi^+>-(-1)^J|\pi^+>|\pi^->)
\]
The summation in (11.6) over $J\lambda$ thus has two parts corresponding to two-pion isospin states $I_C(\pi^- \pi^+)$ and $I_V(\pi^- \pi^+)$ and Generalized Bose-Einstein symmetry conservation and violation, respectively.\\

The two kinds of two-pion states isospin states (12.2) give rise to co-evolution amplitudes that conserve Generalized Bose-Einstein symmetry
\begin{equation}
<J \lambda(\pi^- \pi^+)\eta,I_C(\pi^- \pi^+), \tau_n|<g_ug_d|T|g_ug_d>|0 \tau>=
\end{equation}
\[
\sum \limits_{K \mu} <J \lambda(\pi^- \pi^+),I_C(\pi^- \pi^+), \tau_n|<g_ug_d|T_{fE}|g_ug_d>|K \mu(q \overline{q}),I_C,\tau_{K \mu}>
\]
\[
a_K(m^2)<K \mu(q \overline{q}) \eta, I_C, \tau_{K \mu}|T_i|0 \tau>
\]
and co-evolution amplitudes that violate Generalized Bose-Einstein symmetry
\begin{equation}
<J \lambda(\pi^- \pi^+)\eta,I_V(\pi^- \pi^+), \tau_n|<g_ug_d|T|g_ug_d>|0 \tau>=
\end{equation}
\[
\sum \limits_{K \mu} <J \lambda(\pi^- \pi^+),I_V(\pi^- \pi^+), \tau_n|<g_ug_d|T_{fE}|g_ug_d>|K \mu(q \overline{q}),I_V,\tau_{K \mu}>
\]
\[
a_K(m^2)<K \mu(q \overline{q}) \eta, I_V, \tau_{K \mu}|T_i|0 \tau>
\]
where $I_C,I_V$=0 or 1 for $|I_C(\pi^- \pi^+)>,|I_V(\pi^- \pi^+)>=|S>$ or $|A>$, respectively. Since we focus on resonant modes we have omitted for the sake of brevity the non-resonant contributions with $I_K=2$ on r.h.s. of (12.3) for $J$=even and on r.h.s. of (12.4) for $J$=odd.\\ 

The total co-evolution amplitude is a combination of the two sub-amplitudes (12.3) and (12.4) which in the most general form reads
\begin{equation}
<J \lambda(\pi^- \pi^+)\eta,E_{J \lambda}(\pi^- \pi^+), \tau_n|<g_ug_d|T|g_ug_d>|0 \tau>=
\end{equation}
\[
 \alpha_{J \lambda, \eta}
<J \lambda(\pi^- \pi^+)\eta,I_C(\pi^- \pi^+), \tau_n|<g_ug_d|T|g_ug_d>|0 \tau>+
\]
\[
\omega_{J \lambda, \eta}
<J \lambda(\pi^- \pi^+)\eta,I_V(\pi^- \pi^+), \tau_n|<g_ug_d|T|g_ug_d>|0 \tau> 
\]
where the symbol $E_{J \lambda}(\pi^- \pi^+)$ refers to the fact that the observed two-pion states are in general entangled isospin states
\begin{equation}
<J \lambda(\pi^- \pi^+)\eta,E_{J \lambda}(\pi^- \pi^+)|=
\end{equation}
\[
 \alpha_{J \lambda, \eta}<J \lambda(\pi^- \pi^+)\eta,I_C(\pi^- \pi^+)|
+\omega_{J \lambda, \eta}<J \lambda(\pi^- \pi^+)\eta,I_V(\pi^- \pi^+)| 
\]
The entanglement amplitudes
\begin{eqnarray}
\alpha_{J \lambda, \eta} & = &
<J \lambda(\pi^- \pi^+)\eta,E_{J \lambda}(\pi^- \pi^+)|J \lambda(\pi^- \pi^+)\eta,I_C(\pi^- \pi^+)>\\
\nonumber
\omega_{J \lambda, \eta} & = &
<J \lambda(\pi^- \pi^+)\eta,E_{J \lambda}(\pi^- \pi^+)|J \lambda(\pi^- \pi^+)\eta,I_V(\pi^- \pi^+)>
\nonumber
\end{eqnarray}
are normalized
\begin{equation}
| \alpha_{J \lambda, \eta}|^2+|\omega_{J \lambda, \eta}|^2=1
\end{equation}
To see explicitely the entanglement of the two-pion charge states we rewrite (12.6) in the form
\begin{equation}
<J \lambda(\pi^- \pi^+)\eta,E_{J \lambda}(\pi^- \pi^+)|=
<J \lambda(\pi^- \pi^+)\eta| \Bigl ( a_{J \lambda,\eta}<\pi^- \pi^+)|
+(-1)^J b_{J \lambda,\eta}<\pi^+ \pi^-)| \Bigr )
\end{equation}
where
\begin{eqnarray}
a_{J\lambda,\eta} & = & {1 \over {\sqrt{2}}} \bigl (
\alpha_{J \lambda,\eta}+\omega_{J\lambda,\eta} \bigr )\\
\nonumber
b_{J\lambda,\eta} & = & {1 \over {\sqrt{2}}} \bigl (
\alpha_{J \lambda,\eta}-\omega_{J\lambda,\eta} \bigr )
\nonumber
\end{eqnarray}
For $\alpha_{J \lambda,\eta}=\omega_{J\lambda,\eta}={1\over{\sqrt{2}}}$ the two-pion isospin state is separable.\\

Nucleon helicity amplitudes with definite dipion spin $J$ and helicity $\lambda$ are a combination of nucleon transversity amplitudes with the same spin $J$ and helicity $\lambda$ but with opposite transversity $\tau$~\cite{svec07a,svec07c}. For any combination of solutions of transversity amplitudes the helicity amplitudes must share the same isospin state as the transversity amplitudes. The entanglement amplitudes thus cannot depend on transversity $\tau$ and the quantum numbers $g_\tau$.\\ 

The plane wave co-evolution amplitudes describe the angular distribution of the two-pion states. Their angular expansion in terms of angular amplitudes reads~\cite{svec07a} 
\begin{equation}
<\theta \phi, \eta,E(\pi^-\pi^+), \tau_n|<g_ug_g|T|g_ug_d>|0\tau>=
\end{equation}
\[
\sum \limits_{J,\lambda}Y^J_\lambda(\theta \phi)<J \lambda \eta, E(\pi^-\pi^+), \tau_n|<g_ug_g|T|g_ug_d>|0\tau>\equiv
\]
\[
\sum \limits_{J,\lambda}Y^J_\lambda(\theta \phi)<J \lambda \eta, E_{J\lambda}(\pi^-\pi^+), \tau_n|<g_ug_g|T|g_ug_d>|0\tau> 
\]
where $\theta, \phi$ describe the direction of $\pi^-$ in the two-pion center-of-mass system. Assuming that the final two-pion isospin state $|E(\pi^-\pi^+)>$ is entangled state of symmetric and antisymmetric states
\begin{equation}
|E(\pi^- \pi^+)>=a_S|S>+a_A|A>
\end{equation}
where $|a_S|^2+|a_A|^2=1$, the self-consistency of the angular expansion (12.11) requires 
\begin{eqnarray}
\text {J=even:} & \quad \alpha_{J\lambda, \eta}=a_S, \quad \omega_{J \lambda, \eta}=a_A\\
\nonumber
\text {J=odd: } & \quad \alpha_{J\lambda, \eta}=a_A, \quad \omega_{J \lambda, \eta}=a_S
\end{eqnarray}
We call the entanglement of the state $|E(\pi^-\pi^+)>$ dynamic because it is produced in  a hadron interaction process. In general, the entanglement amplitudes $a_S$ and $a_A$ may depend on energy $s$, dipion mass $m$ and momentum transfer $t$.\\

As the result of the interaction with the environment, there is a change of the entanglement content of the two-pion charge, or isospin, states with definite spin $J$ from maximally entangled states $|S>$ or $|A>$ required by the Generalized Bose-Einstein symmetry with separable $\pi^-\pi^+$ isospin states in the final $\pi^-\pi^+n$ state. The entanglement change comes entirely from the violation of Generalized Bose-Einstein symmetry and demonstrates the fact that the interaction of pion creation process with the environment is a non-dissipative dephasing interaction. Below $\sim$ 1000 MeV where $S$- and $P$-wavea mplitudes dominate it is the environment induced $\rho^0(770)-f_0(980)$ mixing that is responsible for the violations of Generalized Bose-Einstein symmetry and the entanglement of the $\pi^- \pi^+$ isospin states in the final state.\\

In a sequel paper we will show that the requirement of Generalized Bose-Eistein symmetry leads to three relations among partial wave intensities in $\pi^-\pi^+$, $\pi^0 \pi^0$ and $\pi^+ \pi^+$ production for even dipion spins. Available data violate these relations for $S$-wave and $D$-wave intensities providing a direct experimental evidence for a violation of Generalized Bose-Einstein symmetry.\\

In standard Quantum Field Theory the initial and final states of particles are separable Fock states. $CPT$ invariance requires that the amplitudes describing reaction $\pi^- p \to \pi^- \pi^+ n$ also describe the $CPT$ conjugate reaction $\pi^+ \pi^- \overline{n} \to \pi^+ \overline{p}$. This is possible since the initial state $\pi^+ \pi^- \overline{n}$ is separable and thus experimentally preparable. The underlining assumption that makes this statement of $CPT$ invariance possible is that the two reactions are both isolated events in the Universe. As the result, the evolution of the pion creation process is unitary.\\

The evidence for evolution from pure initial states to mixed final states in $\pi^- p \to \pi^- \pi^+ n$ presented in our previous work~\cite{svec07a} necessitates to view the pion creation process as an open quantum system interacting with a quantum environment. According to Wald Theorem~\cite{wald80}, this interaction must violate $CPT$ symmetry. In this work we have reached the conclusion that the interaction with the environment leads to entanglement of $\pi^- \pi^+$ charge, or isospin, states. The entanglement content carried by the produced two-pion states depends on the kinematics and dynamics of the pion creation process and the final states are no longer separable Fock states. The actual final states $<\theta \phi, E(\pi^- \pi^+), \tau_n|<g_ug_d|$ do not posses prepareable $CPT$ conjugate states due to the entanglement of $\pi^-\pi^+$ pairs and because the environment states $|g_ug_d>$ do not have well defined charge conjugate states. As a result the concept of $CPT$ symmetry looses its meaning.\\

The dynamic entanglement of final states is a distinct feature of $CPT$ violation not limited to pion creation processes. Recently Bernab\'{e}u, Mavromatos and Sankar have shown that a $CPT$ violating interaction of free maximally entangled neutral mesons $M^0 \overline{M}^0$ (e.g. $K^0 \overline{K}^0$ or $B^0 \overline{B}^0$ pairs) with an environment of fluctuations of space-time metric (space-time foam) will change the the entanglement of the  $M^0 \overline{M}^0$ pairs and violate Generalized Bose-Einstein symmetry~\cite{bernabeu06a,bernabeu06b}
\begin{eqnarray}
|E(M^0 \overline{M}^0)> & = &
{1 \over{\sqrt{2}}} \bigl ( |\overline{M}^0>|M^0> - |M^0>|\overline{M}^0> \bigr )\\
\nonumber
 & + & {\omega \over{\sqrt{2}}} \bigl ( |\overline{M}^0>|M^0> + |M^0>|\overline{M}^0> \bigr )
\nonumber
\end{eqnarray}
The complex amplitude $\omega$ is a $CPT$ violating parameter associated with the term violating Generalized Bose-Einstein symmetry. It is similar to our amplitude $\omega_{J\lambda, \eta}$ in (12.6). However, $\omega$ arises from $CPT$ violating interaction of the propagating free $M^0 \overline{M}^0$ pairs with environment over macroscopic distances while $\omega_{J\lambda, \eta}$ arises from $CPT$ violating interaction of propagating resonant modes with environment inducing their transitions to two-pion states. Recent measurements of $\omega$ by KLOE Collaboration~\cite{ambrosino06} found small $\omega$ consistent with a zero. Nevertheless, the work of Bernab\'{e}u, Mavromatos and Sankar is the first attempt to explicitely relate entangled final states to quantum gravity. Their work opens the possibility that  $\rho^0(770) - f_0(980)$ mixing observed in the CERN measurements of $\pi^- p \to \pi^- \pi^+ n$ on polarized targets also arises from the low energy manifestations of quantum gravity.

\section{Conclusions.}

The measured $S$- and $P$-wave density matrix elements form an autonomous subspace of reduced density matrix which is analytically solvable at any dipion mass $m$. The presence of $\rho^0(770)$ in the $S$-wave spectra $|S|^2$ arises from the $\rho^0(770)$ peak in the data component $a_1+a_2$ which survives the subtraction of $\rho^0(770)$ peak in the $P$-wave amplitude $3|L|^2$ in the relation $|S|^2=a_1+a_2-3|L|^2$. The presence of $f_0(980)$ in the $P$-wave amplitudes $|L_d|^2$ and $|U_d|^2$ arises from the structures at $f_0(980)$ mass in all other data components. The $\rho^0(770)-f^0(980)$ mixing results from information on dynamics encoded in all measured density matrix elements which also encode the level splitting of the mass spectra. The analytical form of the level splitting of mass spectra reveals the existence of a new quantum number $g$ characterizing the quantum states of the environment and allows to identify the four sets of solutions for the transversity amplitudes with the four co-evolution amplitudes required by the Kraus representation of final state density matrix in $\pi^- p \to \pi^- \pi^+ n$.\\

A model of the $CPT$ violating and non-dissipative interaction of the pion creation process with the environment proposes to explain the $\rho^0(770)-f_0(980)$ mixing and predicts a dynamic entanglement of $\pi^- \pi^+$ pairs in $\pi^- p \to \pi^- \pi+n$. Theoretical work by Wald, Hawking, Ellis, Mavromatos and others suggests that the origin of the environment and its $CPT$ violating non-local interactions is in quantum gravity rendered observable at low energies by the pion creation process. 

\acknowledgments

In 1979 Ludwig van Rossum invited me to join his spin physics group at CEN Saclay, France, to collaborate on the analysis of their measurements of $\pi^+ n \to \pi^+ \pi^- p$ on polarized target at CERN. In 1980 preliminary results suggested evidence for $\sigma(750)$. Ludwig calmed my excitement by pointing out there was no evidence for $\sigma(750)$ in CERN data on $\pi^0 \pi^0$ production but he encouraged me to investigate the evidence for it in $\pi^- \pi^+$ production. Solving the puzzle of $\sigma(750)$ resonance took me a long time and over the many years I often remembered his encouragement. I wish to dedicate this work to Ludwig van Rossum.\\


\begin{thebibliography} {}

\bibitem{erwin61} A.R.~Erwin, R.~March, W.D.~Walker and E.~West, {\sl Evidence for a $\pi \pi$ Resonance in the $I=1$, $J=1$ State}, Phys.Rev.Lett. {\bf 6}, 628 (1961).

\bibitem{hagopian63} V.~Hagopian and W.~Selow, {\sl Experimental Evidence on $\pi \pi$ Scattering Near the $\rho^0$ and $f^0$ Resonances from $\pi^- + p \to \pi + \pi + nucleon$   at 3 BeV/c}, Phys.Rev.Lett. {\bf 10}, 533 (1963).

\bibitem{islam64} M.M.~Islam and R.~Pinn, {\sl Study of $\pi^- \pi^+$ System in $\pi^- + p \to \pi^- + \pi^+ + n$ Reaction}, Phys.Rev.Lett. {\bf 12}, 310 (1964).

\bibitem{patil64} S.H.~Patil, {\sl Analysis of the $S$-wave in $\pi \pi$ Interactions}, Phys.Rev.Lett. {\bf 13}, 261 (1964).

\bibitem{durand65} L.~Durand III and Y.T.~Chiu, {\sl Decay of the $\rho^0$ Meson and the Possible Existence of a $T=0$ Scalar Di-Pion}, Phys.Rev.Lett. {\bf 14}, 329 (1965).

\bibitem{baton65} J.P.~Baton {\sl et al.}, {\sl Single Pion Production in $\pi^-p$ Interactions at 2.75 GeV/c}, Nuovo Cimento {\bf 35}, 713 (1965).

\bibitem{apel72} W.D.~Apel {\sl et al.}, {\sl Results on $\pi \pi$ Interaction in the Reaction $\pi^- p \to \pi^0 \pi^0 n$ at  8 GeV/c}, Phys.Lett. {\bf B41}, 542 (1972).

\bibitem{pennington73} M.R.~Pennington and S.D.~Protopopescu, {\sl How Roy's Equations Resolve Up-Down Ambiguity and Reproduce $S^*$ Resonance}, Phys.Rev. {\bf D7},2591 (1973).

\bibitem{lesquen72} A.~de Lesquen {\sl et al}, {\sl Measurements of Spin Rotation Parameters in Pion-Nucleon Elastic Scattering at 6 and 16 GeV/c}, Phys.Lett. {\bf 40B}, 277 )1972).

\bibitem{cozzika72} G.~Cozzika {\sl et al.}, {\sl The Pion-Nucleon Scattering Amplitudes at 6 and 16 GeV/c}, Phys.Lett. {\bf 40B}, 288 (1972).

\bibitem{lutz78} G.~Lutz and K.~Rybicki, {\sl Nucleon Polarization in the Reaction $\pi^- p \to \pi^- \pi^+ n$}, Max Planck Institute for Physics and Astrophysics, Report MPI-PAE/Exp.El.75, 1978 (unpublished).

\bibitem{becker79a} H.~Becker {\sl et al.}, {\sl Measurement and Analysis of Reaction $\pi^- p \to \rho^0 n$ on Polarized Target}, Nucl.Phys. {\bf B150}, 301 (1979).

\bibitem{becker79b} H.~Becker {\sl et al.}, {\sl A Model Independent Partial-wave Analysis of the $\pi^+ \pi^-$ System Produced at Low Four-momentum Transfer in the Reaction $\pi^- p_{\uparrow} \to \pi^+ \pi^- n$ at 17.2 GeV/c}, Nucl.Phys. {\bf B151}, 46 (1979).

\bibitem{chabaud83} V.~Chabaud {\sl et al.}, {\sl Experimental Indications for a $2^{++}$ non-${\overline q}q$ object},  Nucl.Phys. {\bf B223}, 1 (1983).

\bibitem{rybicki85} K.~Rybicki and I.~Sakrejda, {\sl Indication for a Broad $J^{PC}=2^{++}$ Meson at 840 MeV Produced in the Reaction $\pi^- p \to \pi^- \pi^+ n$ at High $|t|$}, Zeit.Phys. {\bf C28}, 65 (1985).

\bibitem{lesquen85} A.~de Lesquen, L.~van Rossum, M.~Svec {\sl et al.}, {\sl Measurement of the Reaction $\pi^+ n \to \pi^+ \pi^- p$ at 5.98 and 11.85 GeV/c Using a Transversely  Polarized Deuteron Target}, Phys.Rev. {\bf D32}, 4355 (1985).

\bibitem{svec92a} M.~Svec, A.~de Lesquen and L.~ van Rossum, {\sl Amplitude Analysis of Reaction $\pi^+ n \to \pi^+ \pi^- p$ at 5.98 and 11.85 GeV/c}, Phys.Rev. {\bf D45}, 55 (1992).

\bibitem{lesquen89} A.~de Lesquen, L.~van Rossum, M.~Svec {\sl et al.}, {\sl Measurement of the Reaction $K^+ n \to K^+ \pi^- p$ at 5.98 and 11.85 GeV/c Using a Transversely  Polarized Deuteron Target}, Phys.Rev. {\bf D39}, 21 (1989).

\bibitem{svec92b} M.~Svec, A.~de Lesquen and L.~van Rossum, {\sl Amplitude analysis of reaction  $K^+ n _{\uparrow} \to K^+ \pi^- p$ at 5.98 GeV/c}, Phys.Rev. {\bf D45}, 1518 (1992).

\bibitem{alekseev99} ITEP Collaboration, I.G.~Alekseev {\sl et al.}, {\sl Study of the Reaction $\pi^- p \to \pi^- \pi^+ n$ on the Polarized Proton Target at 1.78 GeV/c: Experiment and Amplitude Analysis}, Nucl.Phys. {\bf B541}, 3 (1999).

\bibitem{degroot79b} J.~De ~Groot, {\sl Data Tables of Spin Moments Measured in $\pi^- p \to \pi^- \pi^+ n$ on Polarized Target at 17.2 GeV/c in $\rho^0$ mass region 710-830 MeV at Four-momentum Transfers $-t$=0.000-1.00 $(GeV/c)^2$}, private communication, 1979.

\bibitem{rybicki96} K.~Rybicki, {\sl Data Tables of Spin Moments Measured in $\pi^- p \to \pi^- \pi^+ n$ on Polarized Target at 17.2 GeV/c for Dipiom Masses 580 - 1600 MeV and Four-momentum Transfers $-t$=0.01 - 0.20 $(GeV/c)^2$}, private communication, 1996.

\bibitem{lesquen82} A.~de Lesquen, L.~van Rossum, M.~Svec {\sl et al.}, {\sl Tables des \'{E}l\'{e}ments de Matrice Densit\'{e} de Spin et des Ampltudes de Transversit\'{e} Mesur\'{e}s pour $\pi^+ n \to \pi^+ \pi^- p$ et $K^+ n \to K^+ \pi^- p$ \'{a} 5.98 et 11.85 GeV/c}, Centre d'\'{E}tudes Nucl\'{e}aires de Saclay, Internal Report No. DPhPE 82-01,1982.

\bibitem{donohue79} J.T.~Donohue and Y.~Leroyer, {\sl Is There a Narrow $\sigma$ under $\rho^0$ ?}, Nucl.Phys. {\bf B158}, 123 (1979).

\bibitem{svec84} M.~Svec, in {\sl SPIN 84}, {\sl Observation of a $0^{++}(750)$ Gluonium Candidate in Measurements of $\pi^+ n \to \pi^+ \pi^- p$ on Polarized Target at 5.98 and 11.85 GeV/c}, J.Phys. (Paris) Colloq. {\bf46}, C2 - 281 (1985).

\bibitem{svec92c} M.~Svec, A.~de Lesquen and L.~van Rossum, {\sl Evidence for a Scalar State I=0 $0^{++}(750)$ from Measurements of $\pi N \to \pi^+ \pi^- N$ on a Polarized Target at 5.98, 11.85 and 17.2 GeV/c}, Phys.Rev. {\bf D46}, 949 (1992).

\bibitem{svec96} M.~Svec, {\sl Study of $\sigma(750)$ and $\rho^0(770)$ Production in Measurements of $\pi N \to \pi^+ \pi^- N$ on a Polarized Target at 5.98, 11.85 and 17.2 GeV/c}, Phys.Rev. {\bf D53}, 2343 (1996).

\bibitem{svec97a} M.~Svec, {\sl Mass and Width of the $\sigma(750)$ Scalar Meson from Measurements of $\pi N \to \pi^- \pi^+ N$ on Polarized Target}, Phys.Rev. {\bf D55}, 5727 (1997).

\bibitem{svec02a}  M. ~Svec, {\sl Evidence for a Narrow $\sigma(770)$ Resonance and its Suppression in $\pi \pi$ Scattering from  Measurements of $\pi^- p \to \pi^- \pi^+ n$ on Polarized Target at 17.2 GeV/c}, hep-ph/0210249.

\bibitem{gunter01} BNL E852 Collaboration, J.~Gunter {\sl et al.},
{\sl Partial Wave Analysis of the $\pi^0 \pi^0$ System Produced in $\pi^- p$ Charge Exchange Collisions}, Phys.Rev. {\bf D64}, 072003 (2001).

\bibitem{gale01} O.~Teodorescu, A.K.~Muzunder and Ch.~Gale, {\sl Effects of Meson Mixing on Dilepton Spectra}, Phys.Rev. {\bf C63}, 034903  (2001).

\bibitem{wald80} R.M.~Wald, {\sl Quantum Gravity and Time Irreversibility}, Phys.Rev. {\bf D21}, 2742 (1980).

\bibitem{hawking82} S.H.~Hawking, {\sl The Unpredictability of Quantum Gravity}, Commun.Math.Phys. {\bf 87}, 395 (1982).

\bibitem{hawking84} S.H.~Hawking, {\sl Non-Trivial Topologies in Quantum Gravity}, Nucl.Phys. {\bf B244}, 135 (1984).

\bibitem{ellis84} J.~Ellis, J.S.~Hagelin, D.V.~Nanopoulos and M.~Srednicki, {\sl Search for Violations of Quantum Mechanics}, Nucl.Phys. {\bf B241}, 381 (1984).

\bibitem{huet95} P.~Huet and M.E.~Peskin, {\sl Violations of $CPT$ and Quantum Mechanics in the $K^0 \overline{K}^0$  System}, Nucl.Phys. {\bf B434}, 3 (1995).

\bibitem{ellis96} J.~Ellis, J.L.~Lopez, N.E.~Mavromatos and D.V.~Nanopoulos, {\sl Precission Tests of $CPT$ Symmetry and Quantum Mechanics in the Neutral Kaon System}, Phys.Rev. {\bf D53}, 3846 (1996).

\bibitem{gerber98} H.-J.~Gerber, {\sl Searching for Evolution from Pure States into Mixed States in the Two-State System $K^0 \overline{K}^0$}, Phys.Rev.Lett. {\bf 80}, 0031-9007 (1998).

\bibitem{gerber04} H.-J.~Gerber, {\sl Searching for Evolution from Pure States into Mixed States with Entangled Neutral Kaons}, Eur.Phys.J. {\bf C32}, 229 (2004). 

\bibitem{mavromatos06} N.E.~Mavromatos and S.~Sarkar, {\sl Methods of Approaching Decoherence in the Flavour Sector Due to Space-time Foam}, Phys.Rev. {\bf D74}, 036007 (2006).

\bibitem{bernabeu06a} J.~Bernab\'{e}u, N.E.~Mavromatos and S.~Sarkar, {\sl Decoherence induced $CPT$ violation and Entangled Neutral Mesons}, Phys.Rev. {\bf D74}, 045014 (2006).

\bibitem{bernabeu06b} J.~Bernab\'{e}u, J.~Ellis, N.E.~Mavromatos, D.V.~Nanopoulos and J.~Papavassiliou, {\sl $CPT$ and Quantum Mechanics Tests with Kaons}, hep-ph/0607322.

\bibitem{fidecaro06} M.~Fidecaro and H.-J.~Gerber, {\sl The Fundamental Symmetries in the Neutral Kaon System}, Rep.Prog.Phys. {\bf 69}, 1713 (2006).

\bibitem{ambrosino06} KLOE Collaboration, F.~Ambrosino {\sl et al.}, {\sl First Observation of Quantum Interference in the Process $\phi \to K_S K_L \to \pi^+ \pi^- \pi^+ \pi^-$: Test of Quantum Mechanics and $CPT$ Theorem}, Phys.Lett. {\bf B642}, 315 (2006).

\bibitem{svec07a} M.~Svec, {\sl Evidence for Evolution from Pure States to Mixed States in Pion Creation Process $\pi^- p \to \pi^- \pi^+ n$ on Polarized Target and Its Physical Interpretation}, arXiv:0708.4002 [hep-ph] (2007).

\bibitem{svec07c} M.~Svec, {\sl Determination of $S$- and $P$-Wave Helicity Amplitudes and Non-unitary Evolution in Pion Creation Process $\pi^- p \to \pi^- \pi^+ n$ on Polarized Target}, arXiv:0709.2219 [hep-ph] (2007).

\bibitem{kraus83} K.~Kraus, {\sl States, Effects, and Operations: Fundamental Notions of Quantum Theory}, Lecture Notes in Physics, Vol.190, Springer-Verlag, 1983.

\bibitem{nielsen00} M.A.~Nielsen and I.L.~Chuang, {\sl Quantum Computation and Quantum Information}, Cambridge University Press, 2000.

\bibitem{breuer02} H.P.~Breuer and F.~Petruccione, {\sl The Theory of Open Quantum Systems},
Oxford Univ. Press, 2002.

\bibitem{eadie71} W.T.~Eadie {\sl et al.}, {\sl Statistical Methods in Experimental Physics}, North-Holland Publishing, 1971.

\bibitem{grayer74} G.~Grayer {\sl et al.}, {\sl High Statistics Study of the Reaction $\pi^- p \to \pi^- \pi^+ n$: Apparatus, Method of Analysis, and General Features of Results at 17.2 GeV/c}, Nucl.Phys. {\bf B75}, 189 (1974).

\bibitem{bronshtein54} I.I.~Bronshtein and K.A.~Semendyayev, {\sl Handbook of Mathematics}, Gosudarstvennoe Izdatelstvo Tekhniko-Teoreticheskoi Literatury, Moscow 1954, p.139.

\bibitem{martin70} A.D.~Martin and T.D.~Spearman, {\sl Elementary Particle Theory}, North-Holland, 1970.

\bibitem{bourrely80} C.~Bourrely, E.~Leader and J.~Soffer, {\sl Polarization Phenomena in Hadronic Reactions}, Physics Reports {\bf 59}, 95 (1980).

\bibitem{leader01} E.~Leader, {\sl Spin in Particle Physics}, Cambridge University Press, 2001.

\bibitem{greenberg02} O.W.~Greenberg, {\sl $CPT$ Violation Implies Violation of Lorentz Symmetry}, Phys.Rev.Lett. {\bf 89},231602 (2002). 

\bibitem{gibson76} W.M.~Gibson and B.R.~Pollard, {\sl Symmetry Principles in Elementary Particle Physics}, Cambridge University Press, 1976.

\bibitem{flamm82} D.~Flamm and F.~Sch\"{o}berl, {\sl Introduction to Quark Model of Elementary Particles}, Gordon and Breach Science Publishers, 1982.

\end{thebibliography}
\end{document}